%% file: HIG-21-006_temp.tex
\pdfoutput=1
\documentclass[11pt,twoside,a4paper,cmspaper,final,collab]{cms-tdr}
\def\svnVersion{4d9d070}\def\svnDate{2023/07/15}\def\cmsCernNoTag{CERN-EP-2022-157}\def\cmsCernDate{\today}\def\cmsMessage{Published in the Journal of High Energy Physics as \href{http://dx.doi.org/10.1007/JHEP07(2023)092}{\doi{10.1007/JHEP07(2023)092}.}}
\begin{document}\cmsNoteHeader{HIG-21-006}

\newcommand{\ttH}{\ensuremath{\PQt\PAQt\PH}\xspace}
\newcommand{\fHtt}{\ensuremath{\abs{f_{\CP}^{\PH\PQt\PQt}}}\xspace}
\newcommand{\tH}{\ensuremath{\PQt\PH}\xspace}
\newcommand{\tHq}{\ensuremath{\PQt\PH\PQq}}
\newcommand{\ttW}{\ensuremath{\PQt\PAQt\PW}}
\newcommand{\ttZ}{\ensuremath{\PQt\PAQt\PZ}}
\newcommand{\tHW}{\ensuremath{\PQt\PH\PW}}
\newcommand{\WZ}{\ensuremath{\PW\PZ}}
\newcommand{\ZZ}{\ensuremath{\PZ\PZ}}
\newcommand{\ggH}{\ensuremath{\Pg\Pg\PH}}
\newcommand{\qqH}{\ensuremath{\PQq\PQq\PH}}
\newcommand{\WH}{\ensuremath{\PW\PH}}
\newcommand{\ZH}{\ensuremath{\PZ\PH}}
\newcommand{\ttWW}{\ensuremath{\PQt\PAQt\PW\PW}}
\newcommand{\PHiggs}{\ensuremath{\PH}}
\newcommand{\pp}{\ensuremath{\Pp\Pp}}
\newcommand{\Ptop}{\ensuremath{\PQt}}
\newcommand{\APtop}{\ensuremath{\PAQt}}
\newcommand{\tZ}{\ensuremath{\Ptop\PZ}}
\newcommand{\PZggx}{\ensuremath{\PZ/\gamma^{*}}\xspace}
\newcommand{\Zee}{\ensuremath{\PZggx \to \Pe\Pe}}
\newcommand{\Zmm}{\ensuremath{\PZggx \to \Pgm\Pgm}}
\newcommand{\Ztt}{\ensuremath{\PZggx \to \Pgt\Pgt}}
\newcommand{\Pggx}{\ensuremath{\PGg^{*}}\xspace}
\newcommand{\ttWH}{\ensuremath{\Ptop\APtop\PW\PHiggs}}
\newcommand{\ttZH}{\ensuremath{\Ptop\APtop\PZ\PHiggs}}

\newcommand{\mtop}{\ensuremath{m_{\PQt}}}
\newcommand{\ytop}{\ensuremath{y_{\PQt}}}
\newcommand{\mH}{\ensuremath{m_{\PH}}}

\newcommand{\ckt}{\ensuremath{\kappa_{\PQt}}\xspace}
\newcommand{\ktt}{\ensuremath{\widetilde{\kappa}_{\PQt}}\xspace}
\newcommand{\kv}{\ensuremath{\kappa_{\text{V}}}}
\newcommand{\fourFS}{4FS}
\newcommand{\fiveFS}{5FS}

\renewcommand{\ss}{\ensuremath{\mkern 1mu\mathrm{SS}}}

\newcommand{\twoLeptonssZeroTau}{\ensuremath{2\ell\ss + 0\tauh}}
\newcommand{\twoLeptonssOneTau}{\ensuremath{2\ell\ss + 1\tauh}}
\newcommand{\threeLeptonZeroTau}{\ensuremath{3\ell + 0\tauh}}
\newcommand{\CPBDT}{\ensuremath{\mathrm{BDT}_{\CP}}}

\newcommand{\CP}{\ensuremath{CP}\xspace}

\newlength\cmsTabSkip\setlength{\cmsTabSkip}{1ex}

\providecommand{\cmsTable}[1]{\resizebox{\textwidth}{!}{#1}}

\cmsNoteHeader{HIG-21-006}
\title{Search for \texorpdfstring{\CP}{CP} violation in \texorpdfstring{\ttH}{ttH} and \texorpdfstring{\tH}{tH} production in multilepton channels in proton-proton collisions at \texorpdfstring{$\sqrt{s} = 13\TeV$}{sqrt(s) = 13 TeV}}

\date{\today}

\abstract{
The charge-parity (\CP) structure of the Yukawa interaction between the Higgs (\PH) boson and the top quark is measured in a data sample enriched in the \ttH and \tH associated production, using 138\fbinv of data collected in proton-proton collisions at $\sqrt{s}=13\TeV$ by the CMS experiment at the CERN LHC. The study targets events where the \PH boson decays via $\PH\to\PW\PW$ or $\PH\to\PGt\PGt$ and the top quarks decay via $\PQt\to\PW\PQb$: the \PW bosons decay either leptonically or hadronically, and final states characterized by the presence of at least two leptons are studied. Machine learning techniques are applied to these final states to enhance the separation of \CP-even from \CP-odd scenarios. Two-dimensional confidence regions are set on  \ckt and \ktt, which are respectively defined as the \CP-even and \CP-odd top-Higgs Yukawa coupling modifiers. No significant fractional \CP-odd contributions, parameterized by the quantity \fHtt are observed; the parameter is determined to be $\fHtt = 0.59$ with an interval of $(0.24, 0.81)$ at 68\% confidence level. The results are combined with previous results covering the  $\PH\to\PZ\PZ$ and $\PH\to\PGg\PGg$ decay modes, yielding two- and one-dimensional confidence regions on \ckt and \ktt, while \fHtt is determined to be $\fHtt =  0.28$ with an interval of $\fHtt<0.55$ at 68\% confidence level, in agreement with the standard model \CP-even prediction of $\fHtt=0$.}

\hypersetup{
pdfauthor={CMS Collaboration},
pdftitle={Search for CP violation in ttH and tH production in multilepton channels in proton-proton collisions at sqrt(s) = 13 TeV},
pdfsubject={CMS},
pdfkeywords={CMS, Higgs, Yukawa, CPV}}

\maketitle 

\section{Introduction}
\label{sec:introduction}

The observation of a spin-0 boson by the ATLAS and CMS Collaborations in 2012~\cite{Higgs-Discovery_ATLAS,Higgs-Discovery_CMS,CMS:2019noi} opened a new field of exploration in the realm of particle physics:
the properties of the newly observed boson must be measured in detail to ascertain if it is compatible with the Higgs (\PH{}) boson predicted by the standard model (SM).
Results from the LHC Run~1 and~2 indicate that the coupling of the \PH{} boson to other particles is SM-like~\cite{Khachatryan:2014jba,Sirunyan:2018koj,CMS:2022dwd,ATLAS:2022vkf}.
The SM Yukawa couplings $y_{\textrm{f}}^{\mathrm{SM}}$ of the \PH{} boson to fermions are proportional to the fermion mass $m_{\textrm{f}}$, namely $y_{\textrm{f}}^{\mathrm{SM}} = \sqrt{2}m_{\textrm{f}}/v$, where $v \approx 246\GeV$ denotes the vacuum expectation value of the Higgs field.
The top quark, with a mass of $\mtop = 172.44\pm 0.48\GeV$~\cite{CMS:2015lbj}, is by far the heaviest fermion known to date: its Yukawa coupling $\ytop$ is expected to be of order one.
The large \mtop{} may indicate that it plays a special role in the mechanism of electroweak symmetry breaking~\cite{Dobrescu:1997nm,Chivukula:1998wd,Delepine:1995qs};
deviations of $\ytop$ from the SM prediction would unambiguously indicate the presence of new physics beyond the SM, and thus the determination of $\ytop$ is of special interest in the study of the \PH{} boson.

The associated production of a \PH{} boson with top quark pairs (\ttH{}) has previously been studied by the ATLAS and CMS Collaborations with up to 24.8\fbinv of data recorded at $\sqrt{s} = 7$ and 8\TeV during the LHC Run~1~\cite{CMS_ttH_hbb_multilepton_htt_hgg_8TeV,ATLAS_ttH_hbb1_8TeV,ATLAS_ttH_hbb2_8TeV,ATLAS_ttH_multilepton_with_taus_8TeV,ATLAS_ttH_and_singletopH_hgg_8TeV} and with up to 138\fbinv at $\sqrt{s} = 13\TeV$ during the Run~2~\cite{Sirunyan:2017exp,Aaboud:2017jvq,Aaboud:2017rss,Sirunyan:2018shy,Sirunyan:2018ygk,Sirunyan:2018ouh,Sirunyan:2018mvw,Aad:2020ivc,Sirunyan:2020sum,Sirunyan:2020icl}. The combined analysis of data recorded at $\sqrt{s} = 7$, 8, and 13\TeV allowed for the observation of \ttH{} production with significances of 5.2 and 6.3 standard deviations by the CMS and ATLAS Collaborations, respectively~\cite{Sirunyan:2018hoz,Aaboud:2018urx}. The production of \PH{} bosons in association with a single top quark has been studied by the ATLAS and CMS Collaborations with data recorded during the LHC Run~1~\cite{CMS_singletopH_hbb_hww_htt_hgg_8TeV} and Run~2~\cite{Sirunyan:2018lzm,Aad:2020ivc}. These analyses covered the \PH{} boson decay modes to bottom quark-antiquark pairs, photon pairs, pairs of $\Pgt$ leptons, and combinations of quarks and leptons originating from the decay of intermediate on- or off-shell \PW{} and \PZ{} bosons.

The SM \PH{} boson is even under charge-parity (\CP) inversion; an experimental observation compatible with the presence of a \CP-odd term in the \PH{} boson Lagrangian would be a direct indication of the presence of new physics.  Determining the \CP structure of the couplings of the observed boson is therefore paramount.

The ATLAS and CMS Collaborations have studied the couplings of the \PH{} boson to vector gauge bosons and also tested such interactions for \CP violation~\cite{Chatrchyan:2012jja,Chatrchyan:2013mxa,Khachatryan:2014kca,Khachatryan:2015mma,Khachatryan:2016tnr,Sirunyan:2017tqd,Sirunyan_2019,Aad:2013xqa,Aad:2015mxa,Aad:2016nal,Aaboud:2017oem,Aaboud:2017vzb,Aaboud:2018xdt,Sirunyan:2020sum,Aad:2020ivc,Sirunyan:2021fpv,Sirunyan:2019twz}.
All the studies were compatible with a pure \CP-even \PH{} boson Lagrangian, excluding pure \CP-odd couplings of the \PH{} boson to any of the gauge bosons. However, \CP-violating effects are expected to be theoretically more motivated in \PH{} boson to fermion couplings than those to gauge bosons.
In the couplings to V bosons, \CP-odd contributions enter via nonrenormalizable higher-order operators that are suppressed by powers of $1/\Lambda^{2}$~\cite{Zhang_2011}, where $\Lambda$ is the scale of the physics beyond the SM in an effective field theory, or via nonrenormalizable interaction terms~\cite{Harnik_2013,Ghosh_2019}.
Overall, these \CP-odd contributions are expected to only yield a minor contribution to the coupling.

For the couplings to fermions, a renormalizable \CP-violating coupling may instead occur at tree level.
The top-Higgs Yukawa coupling is therefore a suitable coupling for \CP studies in proton-proton (\pp) collisions~\cite{Gritsan_2016}.
The ATLAS and CMS Collaborations have performed several searches for \CP violation in the coupling of the \PH{} boson to fermions, finding no deviation with respect to the SM prediction of a \CP-even scenario.  The coupling of the \PH{} boson to tau lepton was studied in~\cite{CMS-PAS-HIG-20-006}. Besides, measurements of the \CP structure of the coupling of the \PH{} boson to the top quark were performed using \ttH{} and \tH{} production modes with different final states \eg, two photons~\cite{Sirunyan:2020sum,Aad:2020ivc} or four leptons~\cite{Sirunyan:2021fpv}; the \CP measurements in gluon-gluon fusion production mode can also be interpreted in terms of fermionic couplings~\cite{Sirunyan:2021fpv,HIG-20-007,ATLAS:2021pkb}.

The Yukawa couplings of the \PH{} boson to fermions have been determined with a large uncertainty during the LHC Run~1~\cite{Sirunyan:2018koj}.
During the LHC Run~2, the ratio of $\ytop$ to its SM expectation, $\kappa_\PQt := \ytop/\ytop^{\mathrm{SM}}$,
has been determined by the CMS Collaboration to be within $-0.9 < \kappa_\PQt < -0.7$ or $0.7 < \kappa_\PQt < 1.1$, at 95\% confidence level (\CL)~\cite{Sirunyan:2020icl}
by probing the production of the \PH{} boson in association with one or two top quarks, with subsequent decay to leptonic final states.

\begin{figure}
	\centering
	\includegraphics[width=0.3\textwidth]{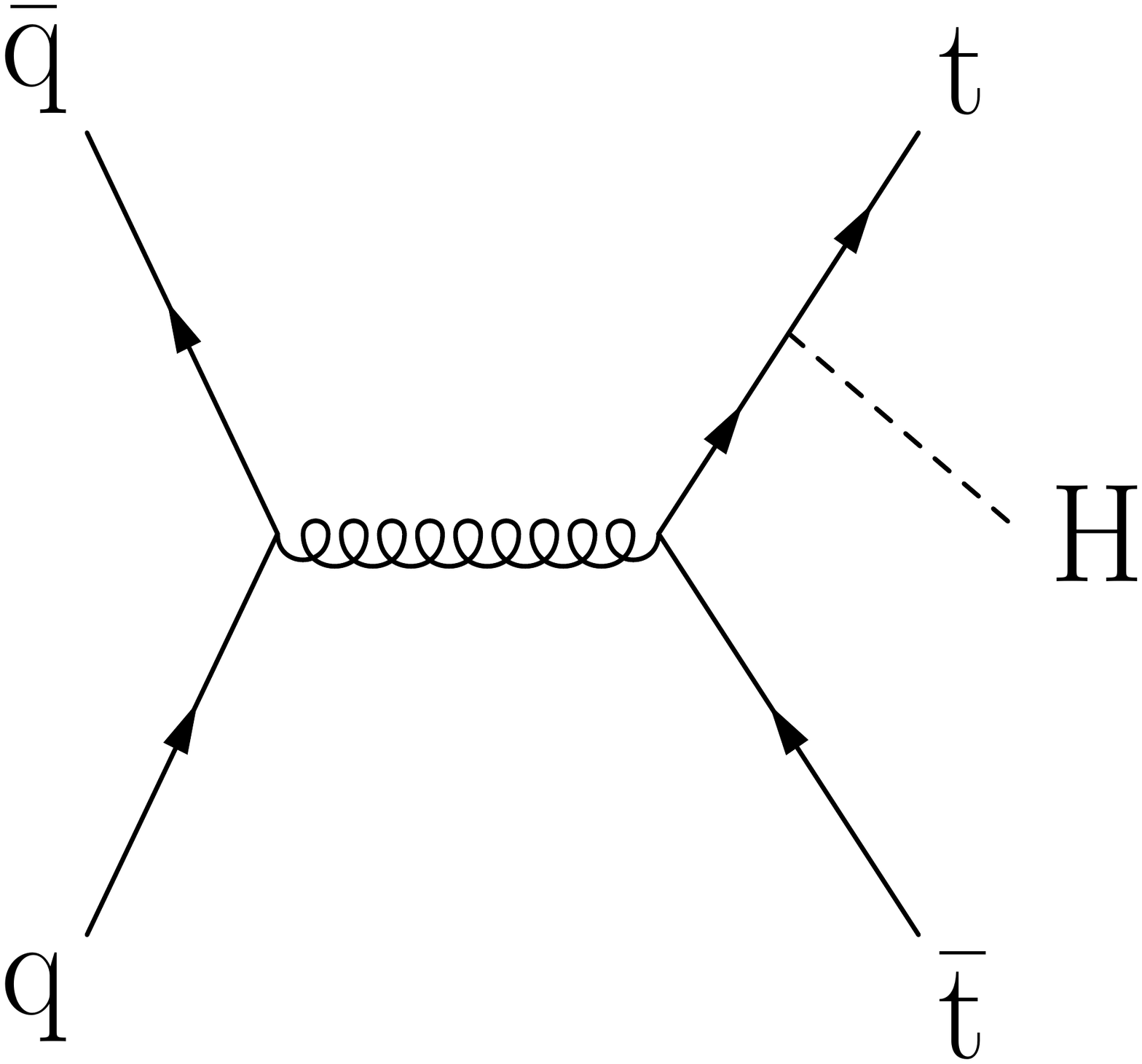} 
	\includegraphics[width=0.3\textwidth]{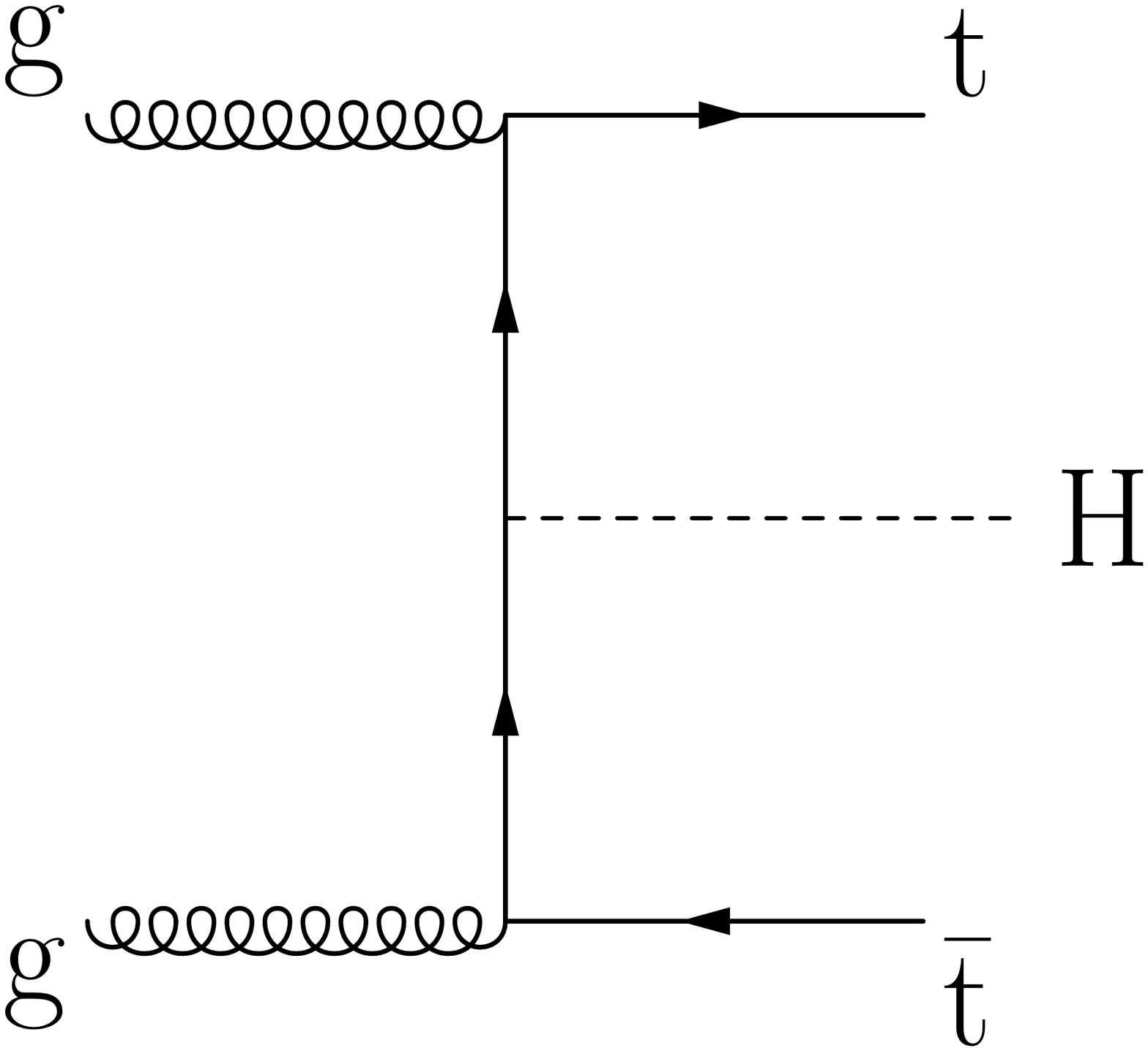}
	\caption{Representative Feynman diagrams for the $\PQt\PAQt\PH$ production processes.}
	\label{fig:FeynmanDiagrams_ttH}
\end{figure}

\begin{figure}
        \centering
                \includegraphics[width=0.25\textwidth]{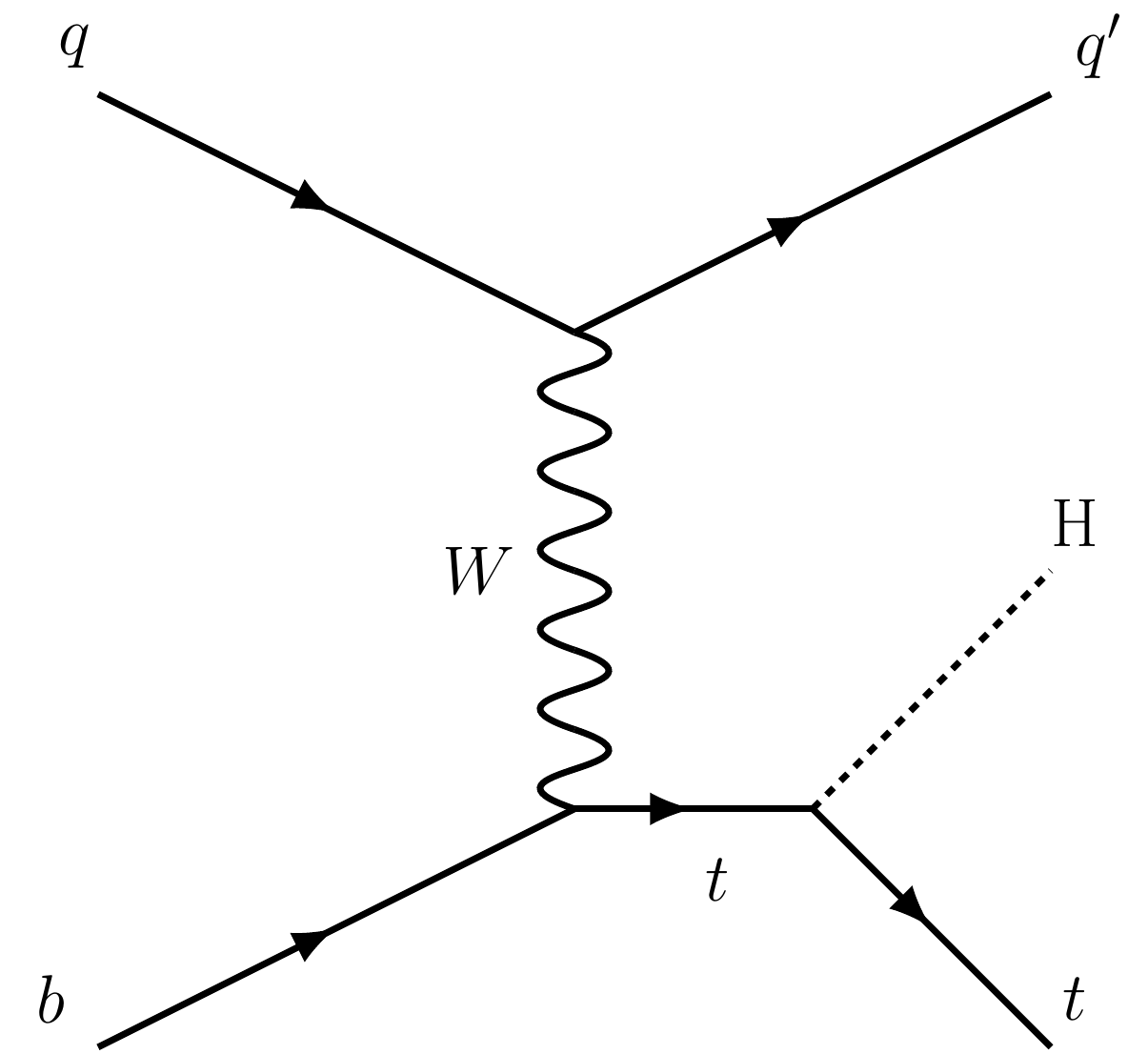}
                \hspace{2cm}
                \includegraphics[width=0.25\textwidth]{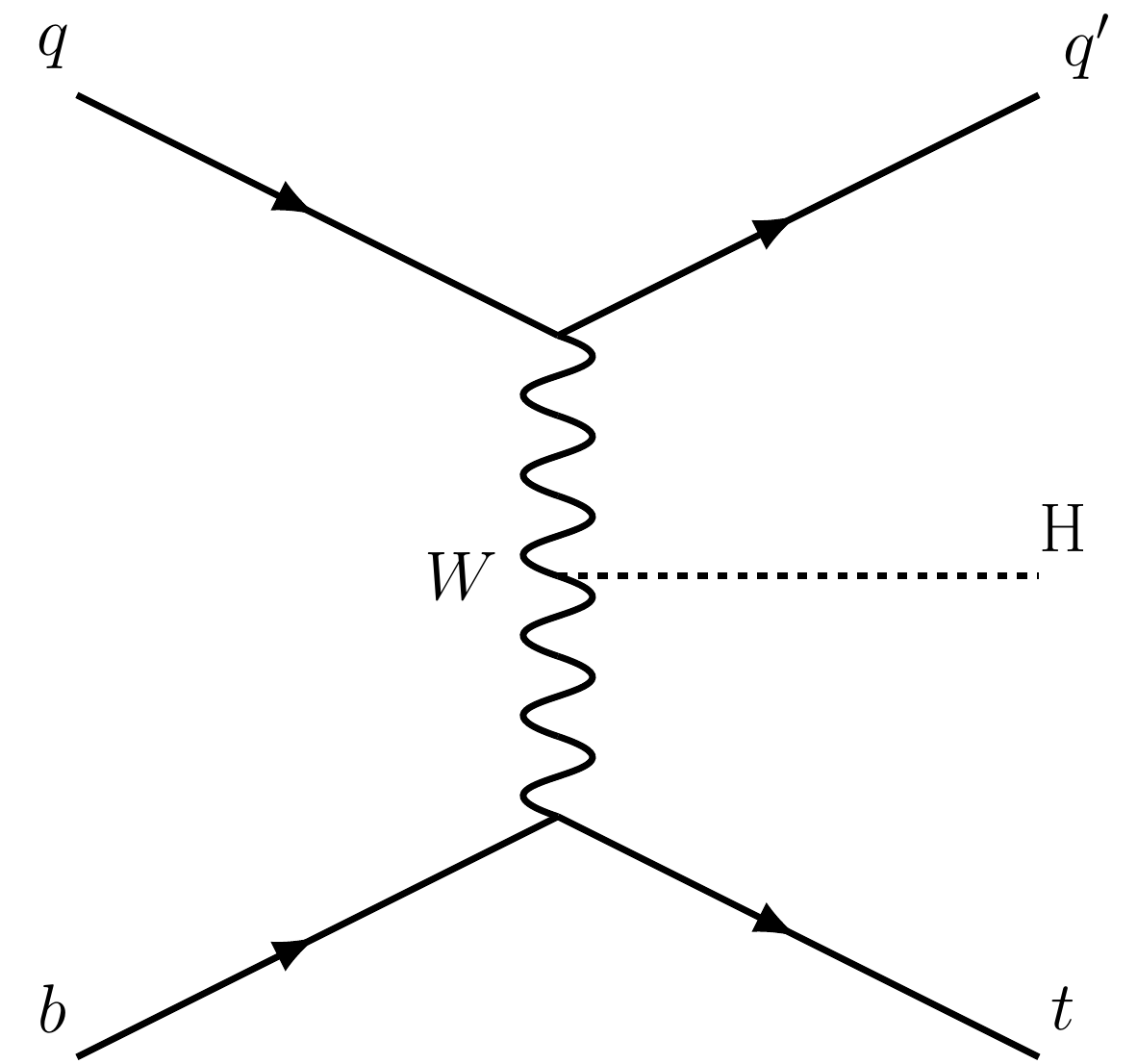}\\
                \vspace{1cm}
                \includegraphics[width=0.25\textwidth]{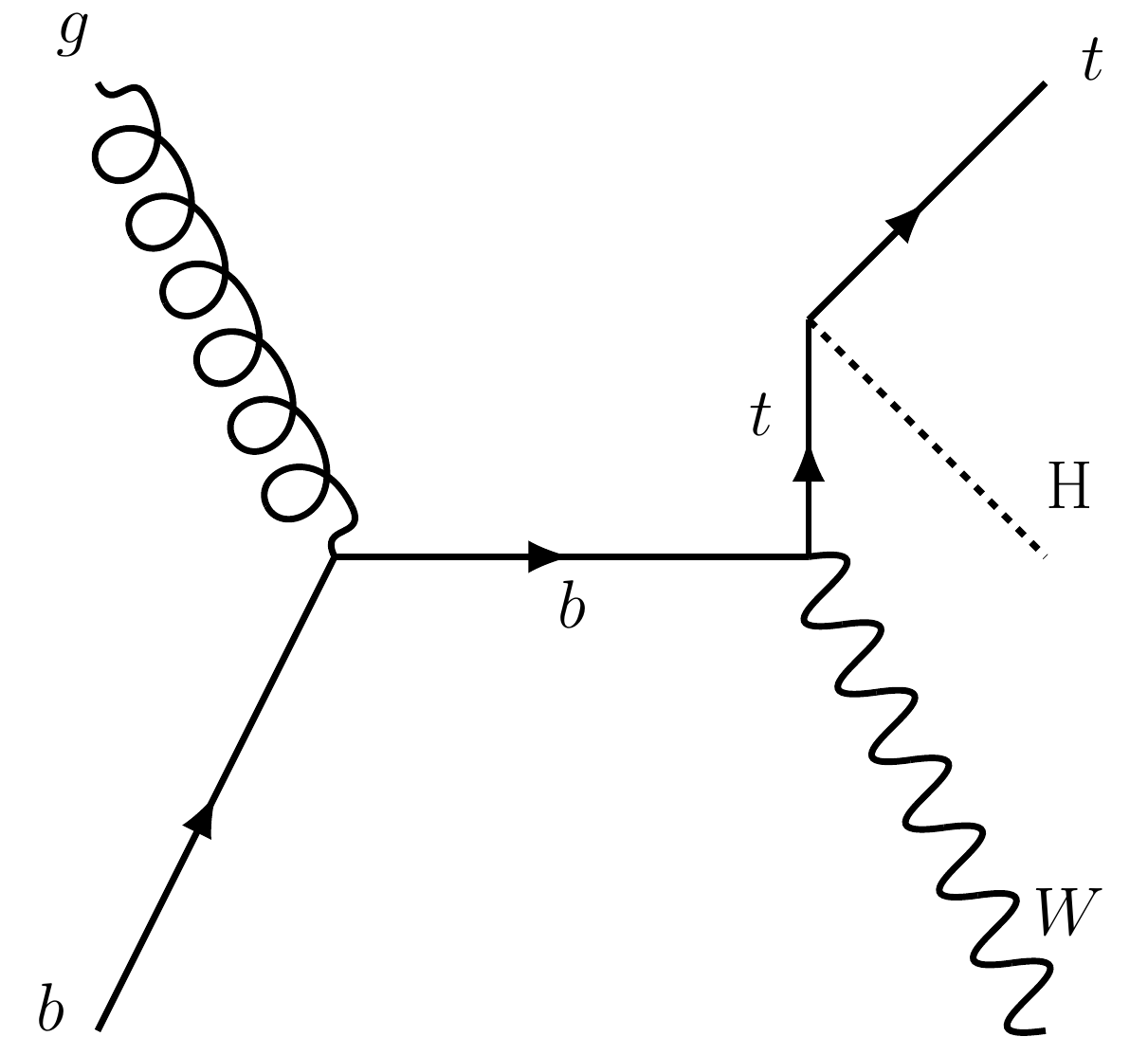}
                \hspace{2cm}
                \includegraphics[width=0.25\textwidth]{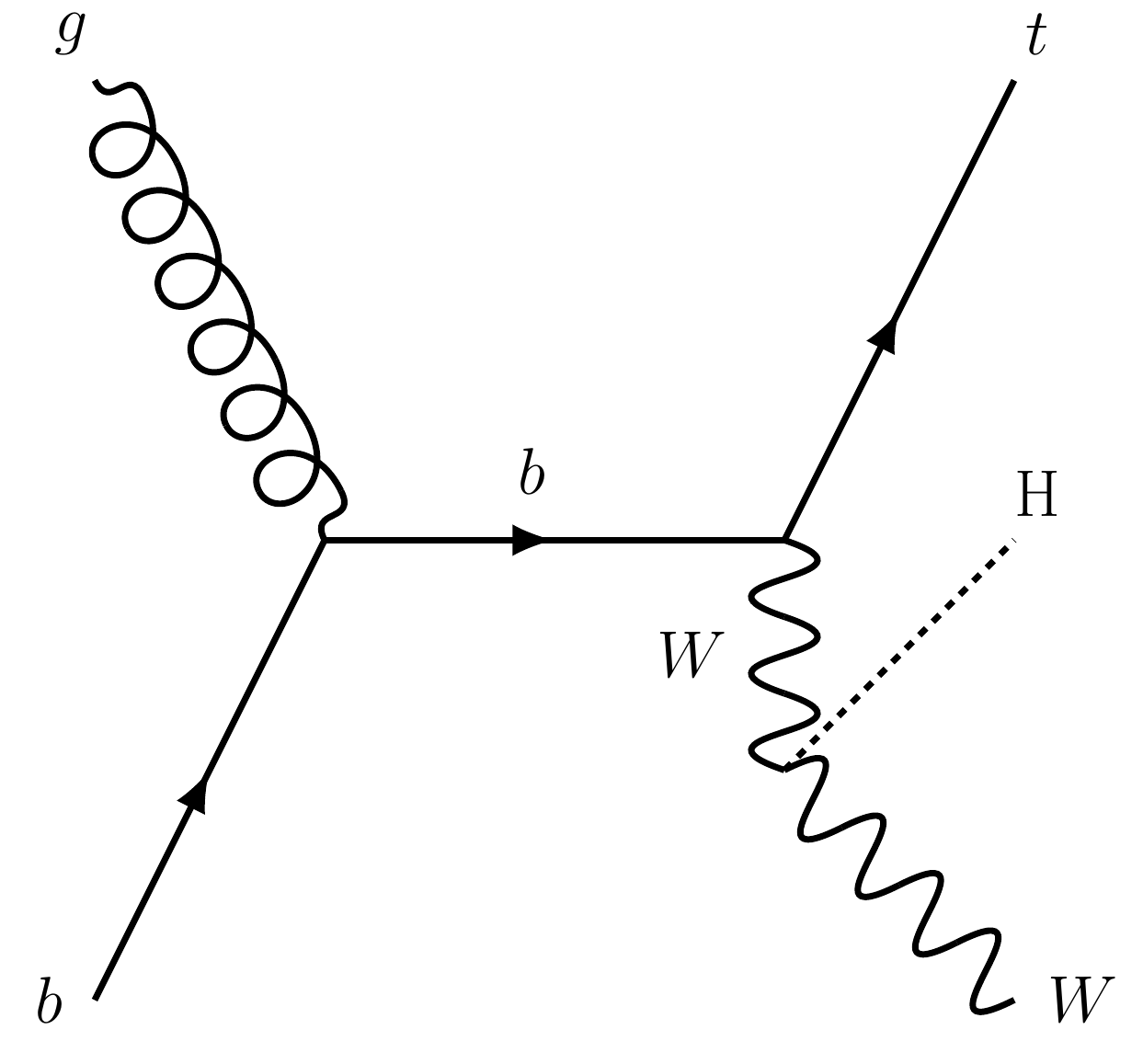}
        \caption{Upper (lower) row: representative Feynman diagrams for the $\PQt\PH$ process in the $t$-channel ($\PQt\PW$-associated) production mode.}
        \label{fig:tH_tchannel}
\end{figure}

The \ttH{} production and the production of a \PH{} boson in association with a single top quark (\tH{}) provide direct access to the magnitude and sign of $\ytop$. The SM cross section for \ttH{} production in \pp{} collisions at $\sqrt{s} = 13\TeV$ amounts to 506.5\unit{fb}~\cite{deFlorian:2016spz} for a \PH{} boson mass of $\mH = 125.09\GeV$~\cite{ATLAS:2015yey}, computed at next-to-leading (NLO) accuracy in quantum chromodynamics (QCD), along with electroweak corrections computed at the same order in perturbation theory. 
The leading-order (LO) Feynman diagrams for \ttH{} production are shown in Fig.~\ref{fig:FeynmanDiagrams_ttH}. The cross section for \tH{} production is significantly smaller, amounting to 74.3\unit{fb}~\cite{deFlorian:2016spz} in the dominant $t$ channel (\tHq{}), computed at NLO accuracy in QCD with the so-called five-flavour scheme (\fiveFS). The \tH{} production also occurs via two additional channels, with lower cross sections, amounting to 15.2\unit{fb}~\cite{deFlorian:2016spz} in the tW-associated channel (\tHW{}) and to 2.9\unit{fb}~\cite{deFlorian:2016spz} in the $s$ channel, both computed at NLO accuracy in QCD with the four-flavour scheme (\fourFS). The contribution of the $s$ channel amounts to 4\% of the contribution of the $t$ channel, and is neglected in this analysis. The diagrams corresponding to the \tHq{} and \tHW{} production modes in the \fiveFS{} can be found in Fig.~\ref{fig:tH_tchannel}. The small rates of the \tHq{} and \tHW{} production processes in the SM are due to the destructive interference between diagrams where the \PH{} boson couples to the top quark (left diagrams of Fig.~\ref{fig:tH_tchannel}) and the corresponding diagrams where the \PH{} boson couples to the \PW{} boson (right diagrams of Fig.~\ref{fig:tH_tchannel}). If $\ytop$ differs from the value predicted by the SM, the cross section for \tH{} production may be modified significantly. The largest enhancement of the \tH{} production rate is attained when $\ytop$= -$\ytop^{\mathrm{SM}}$, referred to as the inverted top coupling (ITC) scenario. In the ITC scenario, the destructive interference between diagrams turns into a constructive interference, and the cross section amounts to 848.0\unit{fb}~\cite{deFlorian:2016spz}. The measurement of the \tH{} production rate thus allows not only a model-independent determination of the $\ytop$ magnitude, but also a determination of its relative sign with respect to the $\PW\PW\PH$ coupling, which makes the study of this process particularly interesting.


In this paper, we report on the measurement of the \CP structure of the \PH{} boson at tree level by studying the \ttH{} and \tH{} processes in final states with multiple electrons, muons, and hadronically decaying tau leptons ($\tauh$). The signatures used are \twoLeptonssZeroTau{}, \twoLeptonssOneTau{}, and \threeLeptonZeroTau{}, which account for the \PH{} boson decay modes $\PH \to \PW\PW$ and $\PH \to \Pgt\Pgt$ (called from now on multilepton decay channel), targeting events in which at least one top quark decays leptonically and providing the highest sensitivity to possible \CP violation effects. The symbol $\ell$ denotes light leptons ($\Pe$, $\Pgm$), and \ss{} means \textit{same-sign}. The measurement is based on data recorded by the CMS experiment in \pp{} collisions at $\sqrt{s} = 13\TeV$ during the LHC Run~2, corresponding to an integrated luminosity of 138\fbinv. We extend the cross section measurement published by the CMS Collaboration~\cite{Sirunyan:2020icl} by interpreting the results in terms of the spin-parity of the \PH{} boson. As in previous analyses~\cite{Sirunyan:2020sum}, the separation of the \ttH{} and \tH{} signals from backgrounds is improved with machine learning techniques, mainly boosted decision trees (BDTs) and artificial deep neural networks (DNNs)~\cite{Kondo:1988yd,Kondo:1991dw}. Machine learning methods are also employed to improve the separation between \CP-odd and \CP-even scenarios, both pure and mixed, for the \ttH{} and \tH{} signals. 
Tabulated results are provided in the HEPData record for this analysis~\cite{hepdata}.

\section{Phenomenology}
\label{sec:pheno}

The Lagrangian for the fermions-Higgs interaction can be written as a superposition of amplitudes with a \CP-even and a \CP-odd phase, where any deviation from the SM values for the couplings would mean \CP violation in the top-Higgs sector and would be described as a beyond-the-SM (BSM) phaenomenon.
Assuming that the spin-0 \PH{} boson is a mass eigenstate, the \ttH{} Lagrangian can be parametererized as follows:
\begin{equation} \label{Lfull}
\mathcal{L}_{\ttH}=\frac{m_{\PQt}}{v} \bar{\psi}_{\PQt} ( \ckt + i\gamma_5 \ktt) \psi_{\PQt} \PH\,,
\end{equation}
where  $\bar{\psi}_{t}$ and $\psi_{t}$ are Dirac spinors, $v$ is the SM Higgs field vacuum expectation value, while \ckt{} and \ktt{} are respectively the \CP-even and \CP-odd top-Higgs Yukawa coupling modifiers.
The parameter $\ckt$ is proportional to  $\mathrm{cos}(\alpha)$, while $\ktt$ is proportional to $\mathrm{sin}(\alpha)$, where $\alpha$ is the mixing angle.
In the SM, there is no \CP violation and therefore $\alpha$ is either $0^{\circ}$ or $180^{\circ}$.
The choice of \ckt{} and \ktt{} affects the coupling and hence the cross section and kinematic properties of both the \ttH{} and \tH{} processes.
We use the variation in the differential cross section of the \ttH{} and \tH{} processes depending on the choice of $\alpha$ derived in Ref.~\cite{Demartin:2015uha}. Based on the choice of $\alpha$, we can broadly identify the three possible scenarios detailed in Table~\ref{tab:CP}.
Kinematic differences between the purely \CP-even, the purely \CP-odd, and the mixed scenario can be exploited to discriminate between them and throw light on the exact \CP scenario that is favoured by Nature.
It is important to note that the cross section of the \ttH{} process is symmetric around $\alpha = 90^{\circ}$ and is therefore not sensitive to the difference between the SM coupling ($\alpha = 0^{\circ}$) and the inverse coupling ($\alpha = 180^{\circ}$).

\begin{table}[h!]
\centering
\topcaption{Possible \CP scenarios}
\begin{tabular}{ l c }
\hline
Scenario & $\alpha$\\
\hline
 Purely \CP-even & $0^{\circ}$ or $180^{\circ}$ \\ 
 Purely \CP-odd & $90^{\circ}$ or $270^{\circ}$ \\ 
 Mixed & $\ne 0^{\circ}, \ne 90^{\circ}, \ne 180^{\circ}$ \\ 
\hline
\end{tabular}
\label{tab:CP}
\end{table}

\section{The CMS detector}
\label{sec:detector}

The central feature of the CMS apparatus is a superconducting solenoid of 6\unit{m} internal diameter, providing a magnetic field of 3.8\unit{T}.
A silicon pixel and strip tracker, 
a lead tungstate crystal electromagnetic calorimeter (ECAL), and a brass and scintillator hadron calorimeter, 
each composed of a barrel and two endcap sections,
are positioned within the solenoid volume.
The silicon tracker measures charged particles within the pseudorapidity range $\abs{\eta} < 2.5$. 
The ECAL is a fine-grained hermetic calorimeter with quasi-projective geometry,
and is segmented into the barrel region of $\abs{\eta} < 1.48$ and in two endcaps that extend up to $\abs{\eta} < 3.0$.
The hadron calorimeter barrel and endcaps similarly cover the region $\abs{\eta} < 3.0$.
Forward calorimeters extend the coverage up to $\abs{\eta} < 5.0$.
Muons are measured and identified in the range $\abs{\eta} < 2.4$ 
by gas-ionization detectors embedded in the steel flux-return yoke outside the solenoid. 
A two-level trigger system~\cite{Khachatryan:2016bia} is used to reduce the rate of recorded
events to a level suitable for data acquisition and storage.
The first level of the CMS trigger system, composed of custom hardware processors, 
uses information from the calorimeters and muon detectors to select the most interesting events with a fixed latency of 4\mus. 
The high-level trigger processor farm further decreases the event rate from around 100\unit{kHz} to about 1\unit{kHz}.
Details of the CMS detector and its performance, together with a definition of the coordinate system and the kinematic variables used in the analysis, are reported in Ref.~\cite{Chatrchyan:2008zzk}.

\section{Data samples and Monte Carlo simulation}
\label{sec:datasamples_and_MonteCarloSimulation}

The data and simulated samples used in this analysis are those used in the inclusive measurement of the \ttH{} and \tH{} cross sections performed by the CMS Collaboration~\cite{Sirunyan:2020icl}.
We report here a few highlights for completeness.

This analysis uses \pp{} collision data recorded at $\sqrt{s} = 13\TeV$ at the LHC during 2016--2018.
Data-taking periods during which the CMS detector was not fully operational are excluded from the analysis.
The total integrated luminosity of the analyzed data set amounts to 138\fbinv,
of which 36.3~\cite{CMS:2021xjt}, 41.5~\cite{CMS-PAS-LUM-17-004}, and 59.8\fbinv~\cite{CMS-PAS-LUM-18-002} have been recorded in 2016, 2017, and 2018, respectively.

The SM cross sections for the \ttH{} and \tH{} signals and for the most relevant background processes estimated from simulation are reported in Table~\ref{tab:backgroundXS}.
The backgrounds arising from nonprompt or misidentified leptons (collectively labelled \textit{nonprompt} in the figures of this paper) and from lepton charge mismeasurement are described in Section~\ref{sec:anal_strategy}; regardless of the processes that originated them, they are fully determined from data. All the other background contributions are determined from simulation.

The \fourFS{} and \fiveFS~\cite{Maltoni:2012pa,Demartin:2015uha} are used to simulate the \tHq{} and \tHW{} processes:
we simulate the \tH{} production at leading order (LO), the \tHq{} process in the \fourFS, and the \tHW{} process in the \fiveFS, so that interference contributions of the latter with \ttH{} production are not present in the simulation.
The contribution from $s$-channel \tH{} production is negligible and is not considered in this analysis.

The \ttH{} and \tH{} signals as well as $\ttbar\Pgg$, $\ttbar\Pggx$, $\tZ$, $\ttWW$, $\PW$+jets, Drell--Yan (DY), $\PW\Pgg$, and $\PZ\Pgg$ backgrounds are simulated at LO using the program \MGvATNLO v2.2.2, v2.3.3 or v2.4.2~\cite{Alwall:2014hca,Alwall:2007fs,Artoisenet:2012st,Frederix:2012ps}. Background arising from \ttbar{} production in association with \PW{} and \PZ{} bosons (\ttW{}, \ttZ{}), 
from triboson ($\PW\PW\PW$, $\PW\PW\PZ$, $\PW\PZ\PZ$, $\PZ\PZ\PZ$, and $\PW\PZ\Pgg$) production, as well as from the production of four top quarks ($\Ptop\APtop\Ptop\APtop$)
are generated at NLO accuracy in perturbative quantum chromodynamics (pQCD) making use of the same program. The modelling of the \ttW{} process includes additional electroweak corrections simulated using \MGvATNLO and proportional to $\alpha_S \alpha_{\text{EW}}^3$~\cite{Frederix:2020jzp,Dror:2015nkp}, where $\alpha_S$ ($\alpha_{\text{EW}}$) is the strong (electroweak) coupling constant. Another sample of \ttH{} events simulated with \MGvATNLO at NLO is used in order to compute one of the uncertainties described in Section~\ref{sec:uncertainties}.  The NLO program \POWHEG v2.0~\cite{Nason:2004rx,Frixione:2007vw,Alioli:2010xd} is used to simulate the backgrounds arising from \ttbar{}, $\Ptop\PW$, and diboson ($\PW^{\pm}\PW^{\mp}$, $\PW\PZ$, and $\PZ\PZ$) production,
from the production of single top quarks,
from SM \PH{} boson production via gluon fusion (\ggH{}) and vector boson fusion (\qqH{}) processes, 
and from the production of SM \PH{} bosons in association with $\PW$ and \PZ{} bosons (\WH{}, \ZH{})
and with \PW{} and \PZ{} bosons along with a pair of top quarks (\ttWH{}, \ttZH{}).

{\tolerance=1200 Parton showering, hadronization, the underlying event as well as the decays of tau leptons, including polarization effects,  are modelled using \PYTHIA~\cite{Sjostrand:2014zea} v8.230 with the tune CP5 ~\cite{Sirunyan:2019dfx}  in 2017 and 2018 Monte Carlo (MC) samples, and \PYTHIA v8.205 with the tunes
CUETP8M1, CUETP8M2, or CUETP8M2T4~\cite{CMS-PAS-TOP-16-021,Khachatryan:2015pea} in 2016 MC samples. 
The matching of matrix elements to parton showers is done using the MLM scheme~\cite{Alwall:2007fs} for the LO samples
and the FxFx scheme~\cite{Frederix:2012ps} for the NLO samples. The \mtop{} assumed in all samples is 172.5\GeV. \par}  

The \ttH{} and \tH{} samples are produced assuming all couplings of the \PH{} boson have the values expected in the ITC scenario. The variation in kinematic properties of \ttH{} and \tH{} signal events is accounted for by applying weights calculated signal event with \MGvATNLO v2.3.3, following the approach suggested in Refs.~\cite{Gainer:2014bta,Mattelaer:2016gcx}. This results in both signal processes to be reweighted by \ckt{} and \ktt{}.
This is different from the inclusive measurement in Ref.~\cite{Sirunyan:2020icl}, where such reweighting was not necessary for the \ttH{} signal,
because any variation of $\ytop$ would only affect the inclusive cross section for \ttH{} production, which increases proportional to $\ytop^{2}$,
leaving the kinematic properties of \ttH{} signal events unaltered.

The presence of simultaneous \pp{} collisions in the same or nearby bunch crossings, referred to as pileup (PU),
is modelled by superimposing inelastic \pp{} interactions, simulated using \PYTHIA, to all MC events.
Simulated events are weighed so that the PU distribution of simulated samples matches the one observed in the data.

All MC events are passed through a detailed simulation of the CMS apparatus, based on \GEANTfour~\cite{Agostinelli:2002hh,Allison:2016lfl},
and are processed using the same version of the CMS event reconstruction software used for the data.

Simulated events are corrected by means of weights or by varying the relevant quantities to account for residual differences between data and simulation.
These differences arise in: trigger efficiencies; reconstruction and identification efficiencies for electrons, muons, and $\tauh$;
the energy scale of $\tauh$ and jets; the efficiency in identifying jets originating from the hadronization of bottom quarks
and the corresponding misidentification rates for light-quark and gluon jets;
and the resolution in missing transverse momentum (\ptmiss).
The corrections are typically at the level of a few percent~\cite{Khachatryan:2010xn,Khachatryan:2016kdb,Sirunyan:2017ezt,Sirunyan:2018pgf,Sirunyan:2019kia}.
They are measured using a variety of SM processes, 
such as DY (\Zee{}, \Zmm{}, and \Ztt{}), \ttbar{}, and $\gamma$+jets production.

\begin{table*}[h!]
  \centering              
  \topcaption{
    Standard model cross sections for the \ttH{} and \tH{} signals as well as for the most relevant background processes estimated from simulation.
    The cross sections are quoted for \pp{} collisions at $\sqrt{s} = 13\TeV$. 
  }
  \label{tab:backgroundXS}
  \resizebox{0.7\textwidth}{!}{
      \begin{tabular}{lc}
      \hline
      Process       & Cross section [fb] \\
      \hline
      \ttH{}        & $507$~\cite{deFlorian:2016spz} \\
      $\tHq$         & $74.3$~\cite{deFlorian:2016spz} \\
      $\tHW$         & $15.2$~\cite{Demartin:2016axk} \\
          [\cmsTabSkip]
          $\ggH$        &   $4.86 \times 10^{4}$~\cite{deFlorian:2016spz} \\
          $\qqH$        &   $3.78 \times 10^{3}$~\cite{deFlorian:2016spz} \\
          $\WH$         &   $1.37 \times 10^{3}$~\cite{deFlorian:2016spz} \\
          $\ZH$         & $884$~\cite{deFlorian:2016spz} \\
          \hline
    \end{tabular}
    \qquad
    \begin{tabular}{lc}
      \hline
      Process       & Cross section [fb] \\
      \hline
      $\ttZ$        & $839$~\cite{deFlorian:2016spz}     \\
      $\ttW$        & $650$~\cite{deFlorian:2016spz,Frederix:2020jzp,Dror:2015nkp}     \\
      $\ttWW$       &      $6.98$~\cite{Alwall:2014hca}\\
          $\PW\PZ$      & $4.50 \times 10^{4}$~\cite{Campbell:2011bn} \\
          $\PZ\PZ$      & $1.69 \times 10^{4}$~\cite{Campbell:2011bn} \\
          \hline
    \end{tabular}
  }
\end{table*}

\section{Event reconstruction and selection}
\label{sec:anal_strategy}

In this Section, we describe the selection criteria that are applied to objects and events in the analysis. This selection
follows closely that of the \ttH{} inclusive analysis in the multilepton channel, which is described in full detail in Ref.~\cite{Sirunyan:2020icl}:
only a brief summary is shown here for completeness.

\subsection{Event reconstruction}

Topologies with prompt leptons are one of the main distinctive features for the signal in this study, where to identify leptons we make use of a dedicated BDT classifier trained to efficiently select
prompt leptons while rejecting leptons coming from other sources, such as
leptons produced in the decay of heavy-flavour hadrons or due to detector mismeasurements.
A less stringent set of lepton requirements (\textit{loose selection}) is used to veto additional leptons in the signal region (SR) and to calculate the misidentification efficiencies associated to the main lepton selection.

Hadronic tau leptons are reconstructed using the hadron-plus-strips algorithm~\cite{Sirunyan:2018pgf}. Similarly to the selection of light leptons and with the same purposes,
different levels of identification are used, based on the output of a deep neural network, as described in Ref.~\cite{Sirunyan:2020icl,CMS:2022prd}. 

Hadronic jets are clustered using candidates compatible with the primary interaction vertex using the anti-\kt{} algorithm~\cite{Cacciari:2008gp}
with a distance parameter $R = 0.4$, and with the particles reconstructed by the particle flow algorithm~\cite{CMS:2017yfk} as inputs. Charged hadrons associated
with PU vertices are excluded from the clustering. The energy of the reconstructed jets is corrected for residual PU effects using the method described in Refs.~\cite{Cacciari:2008gn,Cacciari:2008gp}
and calibrated as a function of jet transverse momentum \pt and $\eta$~\cite{Khachatryan:2016kdb}.

The primary vertex (PV) is taken to be the vertex corresponding to the hardest scattering in the event, evaluated using tracking information alone, as described in Section 9.4.1 of Ref.~\cite{CMS-TDR-15-02}.

To discriminate jets produced by heavy-flavour quarks from those coming from light-flavour quarks and gluons, we make use of the
DeepJet discriminator~\cite{Bols:2020bkb,CMS-DP-2018-058,Sirunyan:2017ezt}.

The missing transverse momentum vector ($\vec{\pt}^{\text{miss}}$) is defined as the negative of the vector \pt sum of all particles reconstructed by the PF algorithm. In this analysis there is a need to discriminate events where the reconstructed \ptmiss arises from resolution effects. In order to do so, a linear combination denoted \textit{linear discriminant} ($\ptmiss LD$) and defined as $0.6\ptmiss + 0.4\mht$ is used, where $\mht$ corresponds to the magnitude of the vector \pt sum of electrons, muons, \tauh, and jets.

\subsection{Event selection and background estimation}

The event selection in this analysis is designed to detect the rare \ttH{} and \tH{} signals in multilepton
final states despite the presence of other SM processes with larger cross section also yielding
leptons in the final state.
Events are collected by a set of single-, double-, and triple-lepton triggers that are designed to maximize the efficiency of the trigger set. The thresholds on the lepton \pt range 12--27 (8--24)\GeV for electrons (muons).
We then target events where the \PH{} boson is produced through the \ttH{}, \tHq{} or \tHW{} process and decays into a pair of \PW{} or \PZ{} bosons or tau leptons.
The \PW{} and \PZ{} bosons can subsequently decay either hadronically or into electrons or muons, while $\PGt$ leptons can decay to electrons, muons, or hadronically (denoted $\tauh$).
To achieve our objectives, we define three event categories depending on lepton and \tauh multiplicity: \twoLeptonssZeroTau{}, \twoLeptonssOneTau{}, and \threeLeptonZeroTau{}.
These regions correspond to the most sensitive ones in the inclusive measurement of \ttH{} and \tH{} production
in the same final state~\cite{Sirunyan:2020icl}: they contain enough \ttH{} events to be sensitive to the signal kinematic properties,
which is a necessary aspect in a search for \CP violation. The selection requirements in these regions are summarized in Table~\ref{tab:selections2}.

Despite the selection mentioned above, the regions are still dominated by background events.
We therefore use multivariate methods, namely artificial neural networks, to separate the contribution
from the various signal and background species.
These discriminators, developed in the context of Ref.~\cite{Sirunyan:2020icl}, take as an input features related to the event kinematics
and provide as an output three (four) output nodes, in the \twoLeptonssOneTau{} and \threeLeptonZeroTau{} (\twoLeptonssZeroTau{}) categories,
that estimate the probability for each event to correspond to \ttH{}, \tHq{}, or any of the backgrounds (to \ttH{}, \tHq{}, \ttW{}, or any
other background). An extra node is added in the \twoLeptonssZeroTau{} category to gain sensitivity to the
irreducible \ttW{} background.
Events are classified according to their most probable node, with an additional categorization in lepton charge and medium \PQb-tagged jet multiplicity
in the \twoLeptonssZeroTau{} and \threeLeptonZeroTau{}. We define as \textit{medium \PQb-tagged} a jet whose value of the DeepJet discriminator is larger than a \textit{medium working point} that corresponds to a misidentification probability of 1\% for light-quark and gluon jets.

Kinematic differences between the \ttH{} \CP-even component and a potential \ttH{} \CP-odd component are then exploited by means of an additional multivariate discriminator.
A dedicated gradient-BDT~\cite{Hastie-Tibshirani-Friedman,Breiman-Friedman-Olshen-Stone} that we label \CPBDT{}, is trained for each category to distinguish
between these two components, as described in Section~\ref{sec:cp_discrimination}.
Other families of discriminators have been tested, including a likelihood ratio regression based on the methods proposed by Ref.~\cite{Brehmer2018}, finding no significant improvements with respect to the main BDT approach.
The details of the \CP hypothesis discrimination are described in detail in Section~\ref{sec:cp_discrimination}.

In order to determine the normalization of the \ttZ, \WZ{}, and \ZZ{} processes, we make use
of control regions (CRs) enriched in these processes. Two CRs are defined, requiring three and four leptons in the final state.
The three-lepton CR is defined by inverting the \PZ boson veto present in the \threeLeptonZeroTau{} SR. The selection criteria on jets and \PQb-tagged jets are dropped in this region. Events are then classified according to the number of jets and \PQb-tagged jets, in order to be sensitive to \WZ{} and \ttZ{} production.
A four-lepton CR is defined by events with four leptons, two of which are taken to form a \PZ boson candidate. Events in this region are classified according to the number of \PZ boson candidates and the jet and \PQb-tagged jet multiplicity.

The contribution of signal and background events is estimated based on methods outlined in Ref.~\cite{Sirunyan:2020icl}.
Backgrounds are classified as either reducible or irreducible depending on the source of the reconstructed leptons passing the object selection criteria.
A background is considered reducible when one or more of the reconstructed leptons passing the main object selection criteria do not correspond to a lepton originated in the prompt decay of a \PW or \PZ boson or a tau lepton decay.
The background is considered irreducible otherwise. 

We distinguish three sources of reducible background contributions: misidentified leptons and $\tauh$s (\textit{misidentified lepton}), conversions of a photon into an electron-positron pair in the tracker material (\textit{conversions}), and mismeasurement of the lepton charge (\textit{charge flip}). The main contribution to misidentified lepton background stems from \ttbar{} production, reflecting the large cross section for this background process. The conversions background is typically caused by $\ttbar\gamma$ events in which one electron or positron produced in the photon conversion carries most of the energy of the converted photon, whereas the other electron or positron is of low energy and fails to get reconstructed. We refer to such photon conversions as \textit{asymmetric conversions}. The contribution of this background is heavily suppressed by requiring electron tracks not to have
missing hits in the innermost layers of the silicon tracker and by vetoing candidates that are matched to a reconstructed conversion vertex. The flips background is specific to the \twoLeptonssZeroTau{} and \twoLeptonssOneTau{} channels and consists in events where the charge of a reconstructed lepton is mismeasured.

The mismeasurement of the electron charge typically results from the emission of a hard bremsstrahlung photon,
followed by an asymmetric conversion of this photon.
The reconstructed particle is typically the electron or positron that carries most of the energy of the converted photon,
resulting in an equal probability for the reconstructed particle to have either the same or opposite charge
compared to the charge of the electron or positron that emitted the bremsstrahlung photon~\cite{Khachatryan:2015hwa}.
The main contribution to the flips background stems from \ttbar{} events in which both top quarks decay semi-leptonically. In case of the \twoLeptonssOneTau{} channel, a quark or gluon jet is additionally misidentified as $\tauh$.
The probability of mismeasuring the charge of muons is negligible in this analysis. 

The misidentified leptons and flips backgrounds are largely suppressed by the lepton and \tauh selection described in Section~\ref{sec:anal_strategy}, and are estimated using control samples in data with techniques based on the misidentification
probability method~\cite{Sirunyan:2018shy}, whereas the conversions background is modelled using the MC simulation. More details can be found in Ref.~\cite{Sirunyan:2020icl}.

Irreducible backgrounds are estimated using samples of simulated events. We call \textit{rare} backgrounds to the aggregate of background processes which typically yield a minor contribution 
and include such processes as \tZ{} production,
the production of \ss{} \PW{} boson pairs, triboson, and $\Ptop\APtop\Ptop\APtop$ production. In particular we include under this label all backgrounds estimated from simulation
other than those arising from \ttW{}, \ttZ{}, and diboson backgrounds or from SM \PH{} boson production via the processes \ggH{}, \qqH{}, \WH{}, \ZH{}, \ttWH{}, and \ttZH{}. 

Additional details on the background categorization and estimation are given in Ref.~\cite{Sirunyan:2020icl}.

\begin{table}[h!t]
\topcaption{
  Event selections applied in the \twoLeptonssZeroTau{}, \twoLeptonssOneTau{}, and \threeLeptonZeroTau{} categories.
}
\label{tab:selections2}
{\centering

  \resizebox{\textwidth}{!}{
\begin{tabular}{lccc}
\hline
Selection & \twoLeptonssZeroTau{} & \twoLeptonssOneTau{} & \threeLeptonZeroTau{} \\
\hline
Trigger & \multicolumn{2}{c}{Single- and double-lepton triggers} & Single-, double- and triple-lepton triggers \\
[\cmsTabSkip]
 Lepton \pt & $\pt > 25$ / $15\GeV$ & $\pt > 25$ / $15$ ($\Pe$) or $10\GeV$ ($\Pgm$)  & $\pt > 25$ / $15$ / $10\GeV$\\
 Lepton $\eta$ & \multicolumn{2}{c}{$\abs{\eta} < 2.5$ ($\Pe$) or $2.4$ ($\Pgm$)}  & $\abs{\eta} < 2.5$ ($\Pe$) or $2.4$ ($\Pgm$)\\
 $\tauh$ \pt & \NA & $\pt > 20\GeV$ & \NA \\
$\tauh$ $\eta$ & \NA & $\abs{\eta} < 2.3$ & \NA\\
Charge requirements & \multicolumn{2}{c}{2 same-sign leptons and}& \multirow{2}{*}{\NA} \\
 & \multicolumn{2}{c}{charge quality requirements} & \\
 & & $\sum\limits_{\ell,\tauh} q = \pm 1$ & $\sum\limits_{\ell} q = \pm 1$ \\
[\cmsTabSkip]
 Jet multiplicity$^{**}$ &  \multicolumn{2}{c}{$\geq$3 jets} & $\geq$2 jets \\
 $\PQb$ tagging requirements$^{**}$ & \multicolumn{3}{c}{$\geq$1 medium \PQb-tagged jet or $\geq$2 loose \PQb-tagged jets} \\
 Light jet and \PQb{} tag$^{***}$ & \multicolumn{3}{c}{$\geq$1 light jets, $\geq$1 medium \PQb-tagged jet} \\
 [\cmsTabSkip]
 Missing transverse momentum & \multicolumn{2}{c}{$\ptmiss LD > 30\GeV\,^{\mathsection}\,$} & $\ptmiss LD > 45\GeV\,^{\ddagger}\,^{**}\}$ \\
[\cmsTabSkip]
 Dilepton mass & \multicolumn{3}{c}{$m_{\ell\ell} > 12\GeV\,^{*}$, $\abs{m_{\ell\ell} - m_{\PZ}} > 10\GeV$ } \\
 \hline
\end{tabular}
}}
\footnotesize{$^{*}$ Applied on all pairs of leptons that pass loose selection.} \\
\footnotesize{$^{**}$ If events do not pass these thresholds, $^{***}$ is required.} \\
\footnotesize{$^{\ddagger}$ If the event contains a same-flavour opposite-sign (SFOS) lepton pair and $N_{\text{jet}} \leq 3$. If $N_{\text{jet}} \leq 3$ but there is no SFOS lepton pair, the threshold is 30\GeV.} \\
\footnotesize{$^{\mathsection}$ Applied only to $\Pe\Pe$ events.} \\
\end{table}

\section{The \texorpdfstring{\CP}{CP} discrimination}
\label{sec:cp_discrimination}

In order to separate \CP-even and \CP-odd scenarios, we trained three separate BDT classifiers, one for each of the channels: \twoLeptonssZeroTau{}, \twoLeptonssOneTau{}, and \threeLeptonZeroTau{}. All are based on the \textsc{XGBoost} gradient boosting implementation~\cite{DBLP:journals/corr/ChenG16} and 50\% of the samples are used to train the BDTs, 40\% for testing, and 10\% as validation data set during training.
The hyperparameters of each classifier are optimized using a sequential grid search.

Each of the channel-specific BDTs used a different set of input features. The transverse momentum, $\eta$, and $\phi$ of the final-state particles are used as input variables in all BDTs. The jet three-vectors are directly used only in the \threeLeptonZeroTau{} category. Other variables, such as the momentum vectors of the four leading jets, have also been tested in the other final states. We found that they don't provide any increase in performance with respect to the current set, and their removal did not result in any performance loss. Other kinematic quantities such as the distance in the ($\eta$, $\phi$) plane between the final-state objects are also used in all three categories. 

The invariant mass of the system composed by the leptons in the final state (including $\PGt$s), the \ptmiss, and an appropriate number of jets (depending on the final state), proved to be a useful variable to discriminate between \CP-even and \CP-odd states. We call this variable invariant mass of the reconstructed \ttH{} and is defined as:
\begin{equation}\label{eq:mtth}
    M_{\ttH{}} = \| \sum_i p^{\text{lep}_i}+ (\pt^{\text{miss}}, \vec{\pt}^{\text{miss}}) +  \sum_{i \leq k} p^{\text{jet}_i*}\|\;,
\end{equation}
where $\|\cdot\|$ denotes the norm in the Minkowski space and $p^{\text{lep}_i}$ ($p^{\text{jet}_i*}$) are the four-momenta of the leptons (jets). The number $k$ of jets used to compute the invariant mass of the system is taken accordingly to the expected number of jets given by the \ttH{} subsequent decays in the final state corresponding to the category, \ie, $k = 6$ (4) in the final state \twoLeptonssZeroTau{} (\twoLeptonssOneTau{} and \threeLeptonZeroTau{}). If more than the expected number of jets are present in the event, because of initial- or final-state radiation or because of misreconstructed jets from detector noise or PU, then the appropriate subset of jets is chosen by highest \pt. This mass variable is used in all three BDTs.

For the \twoLeptonssZeroTau{} final state, the transverse masses of each lepton and \ptmiss are included in the training. We define the transverse mass as:
\begin{equation}\label{eq:mt}
    m_{\mathrm{T}}(l_i,\ptmiss) = \sqrt{ \left(\abs{\vec{p}_{\mathrm{T,l_i}}}+\abs{\vec{\pt}^{\text{miss}}} \right)^2 - \left\abs{ \vec{p}_{\mathrm{T,l_i}}+\vec{\pt}^{\text{miss}} \right}^2}\;,    
\end{equation}
where $\vec{p}_{\mathrm{T,l_i}}$ is the transverse momenta of the corresponding lepton.

The jet multiplicity, the $\ptmiss$, the $\phi$ component of the $\ptvecmiss$ vector, and the average distance in the ($\eta$, $\phi$) plane between the jets are also included, to further increase the performance in the \twoLeptonssZeroTau{} and \twoLeptonssOneTau{} categories. 

Besides, jets coming from the hadronic decay of the top quark can also be exploited to enhance the discrimination between the two \CP scenarios.  A BDT is computed over a given triplet of jets, allowing to identify the triplets of jets produced in the hadronic decay of a top quark, as described in Ref.~\cite{Sirunyan:2020icl}. 
We use the highest score of all possible jet triplets present in the event as input variable in the \twoLeptonssZeroTau{} and \twoLeptonssOneTau{} categories. For the \twoLeptonssOneTau{}, the output of a BDT classifier that identifies jets from \PW boson decays from the $\PH\to\PW\PW$ decay chain is also used~\cite{Sirunyan:2020icl}.

\begin{figure}
\centering
\includegraphics*[height=61mm]{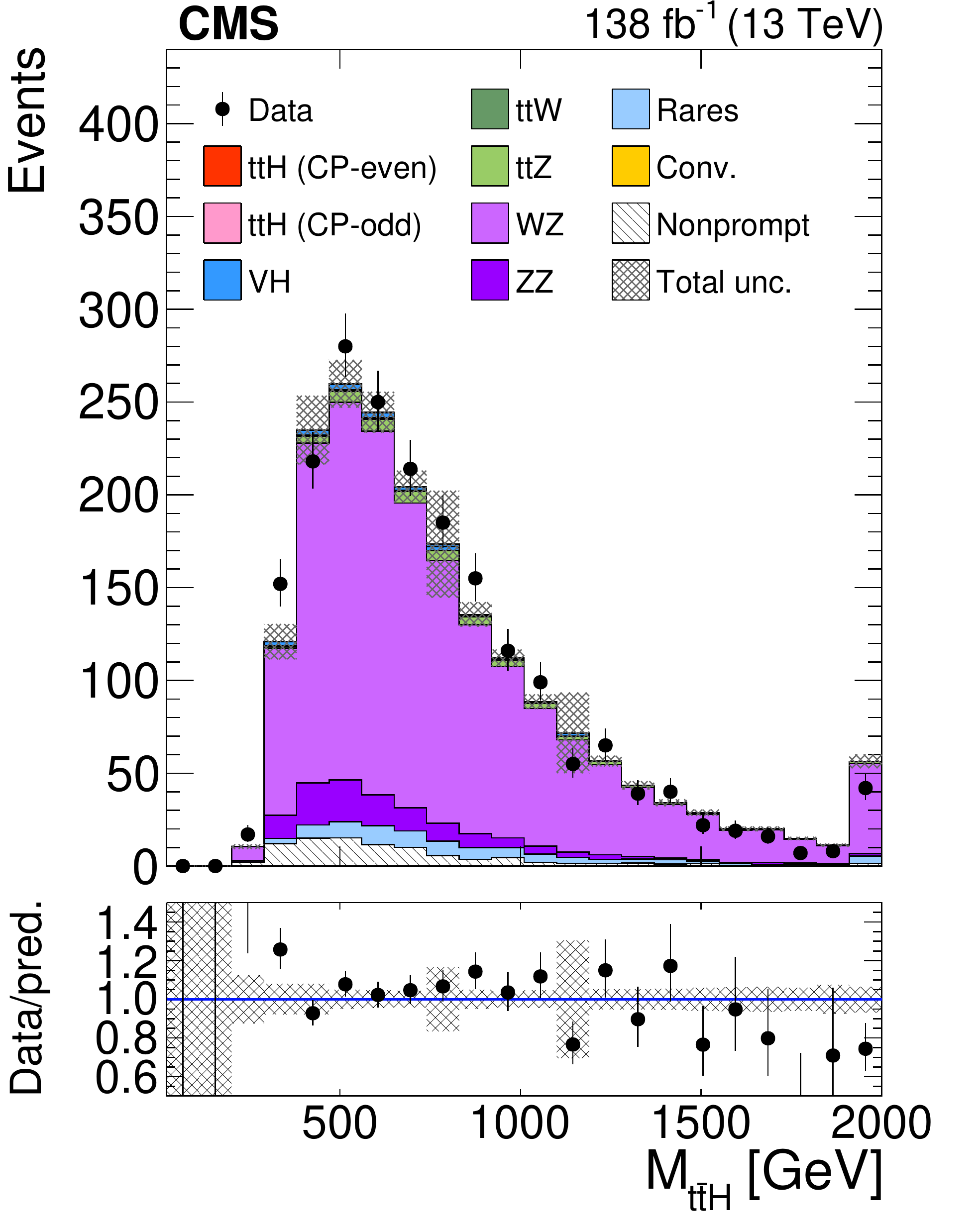}
\includegraphics*[height=61mm]{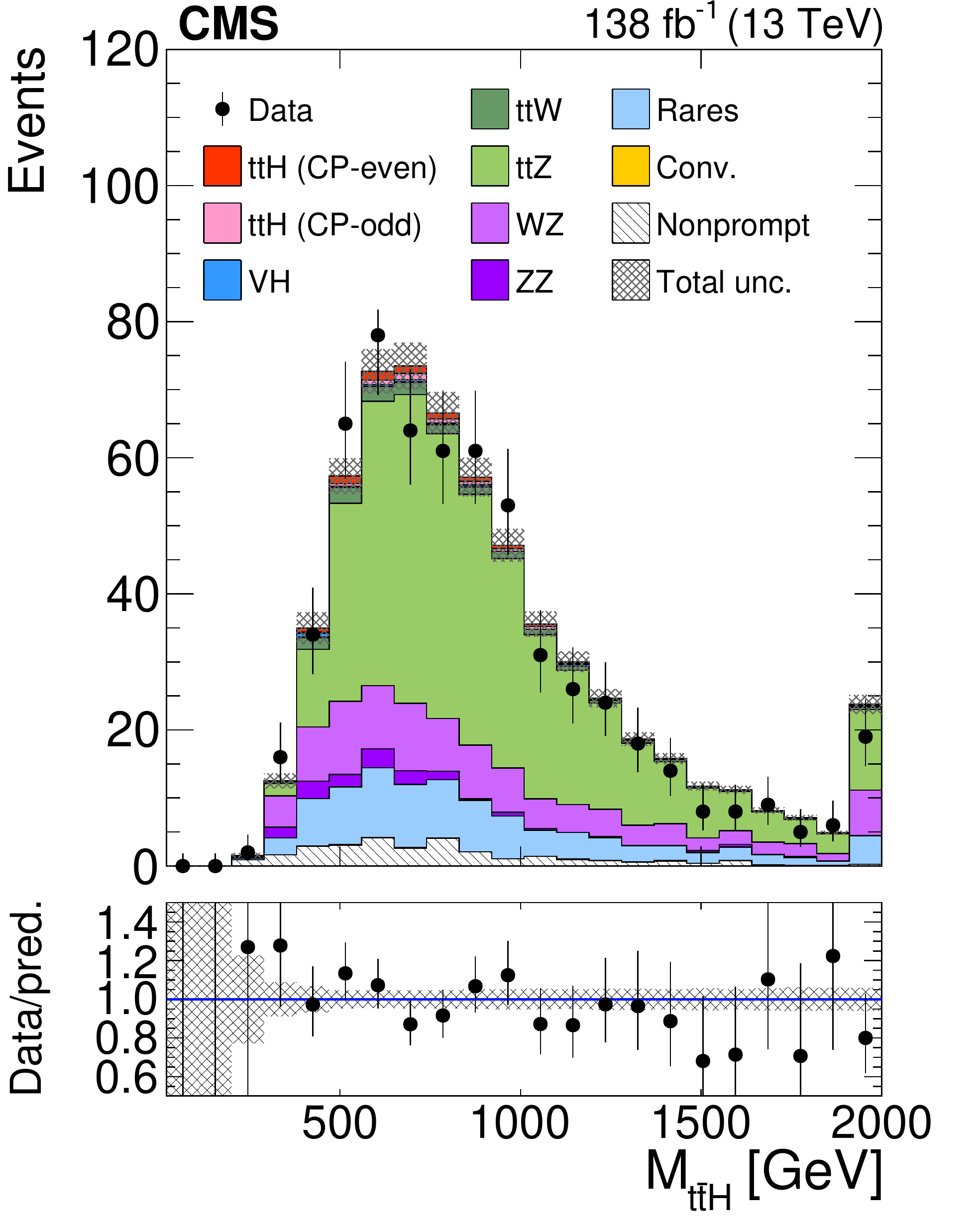}
\includegraphics*[height=61mm]{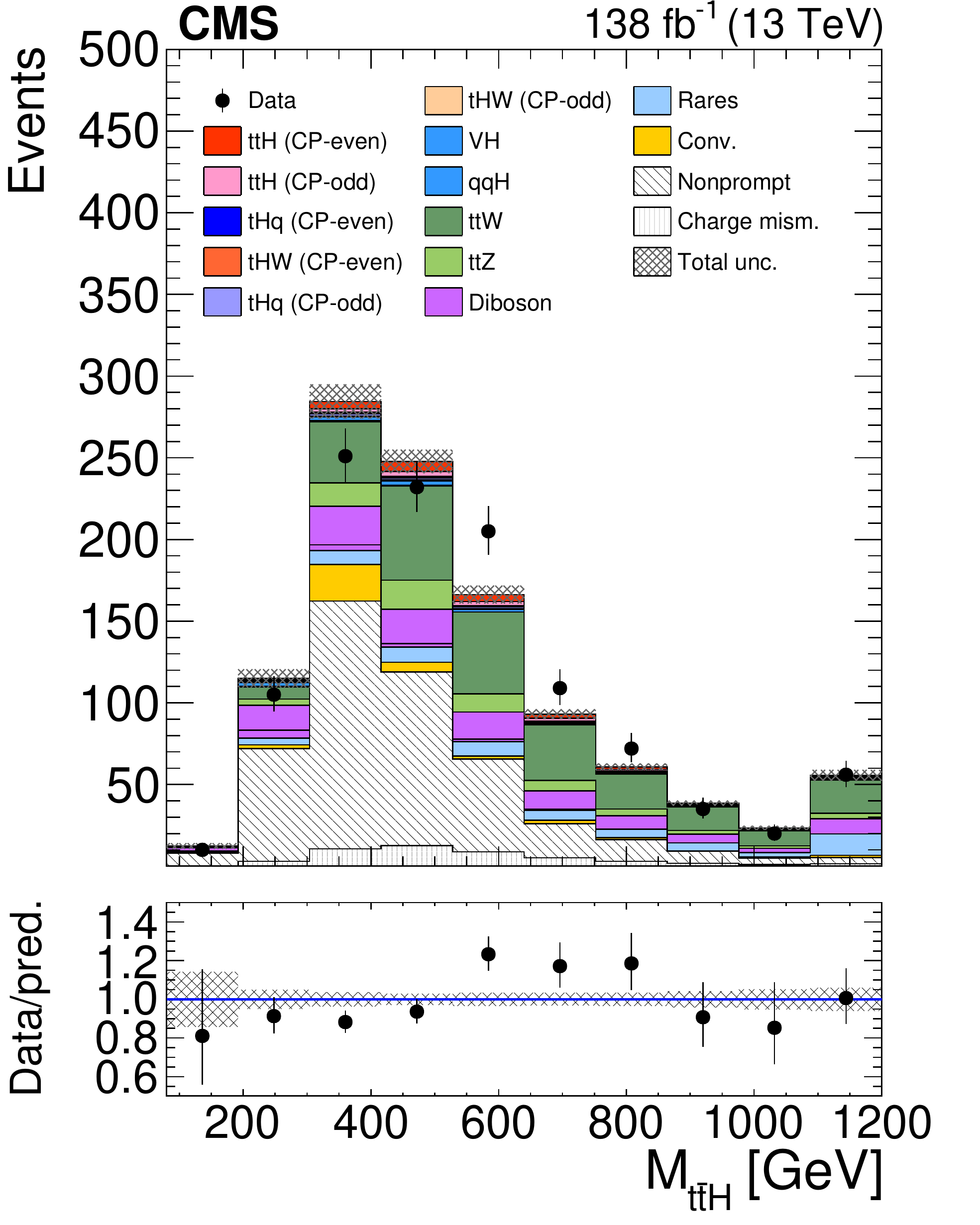}
\caption{
One of the most important input variables for the \CP discriminant, $M_{\ttH{}}$, in the three validation regions enriched from left to right in $\PW\PZ$, $\PQt\PAQt\PZ$ and  misidentified lepton background.  
}
\label{validationplots1}
\end{figure}

\begin{figure}
\centering
\includegraphics*[height=60mm]{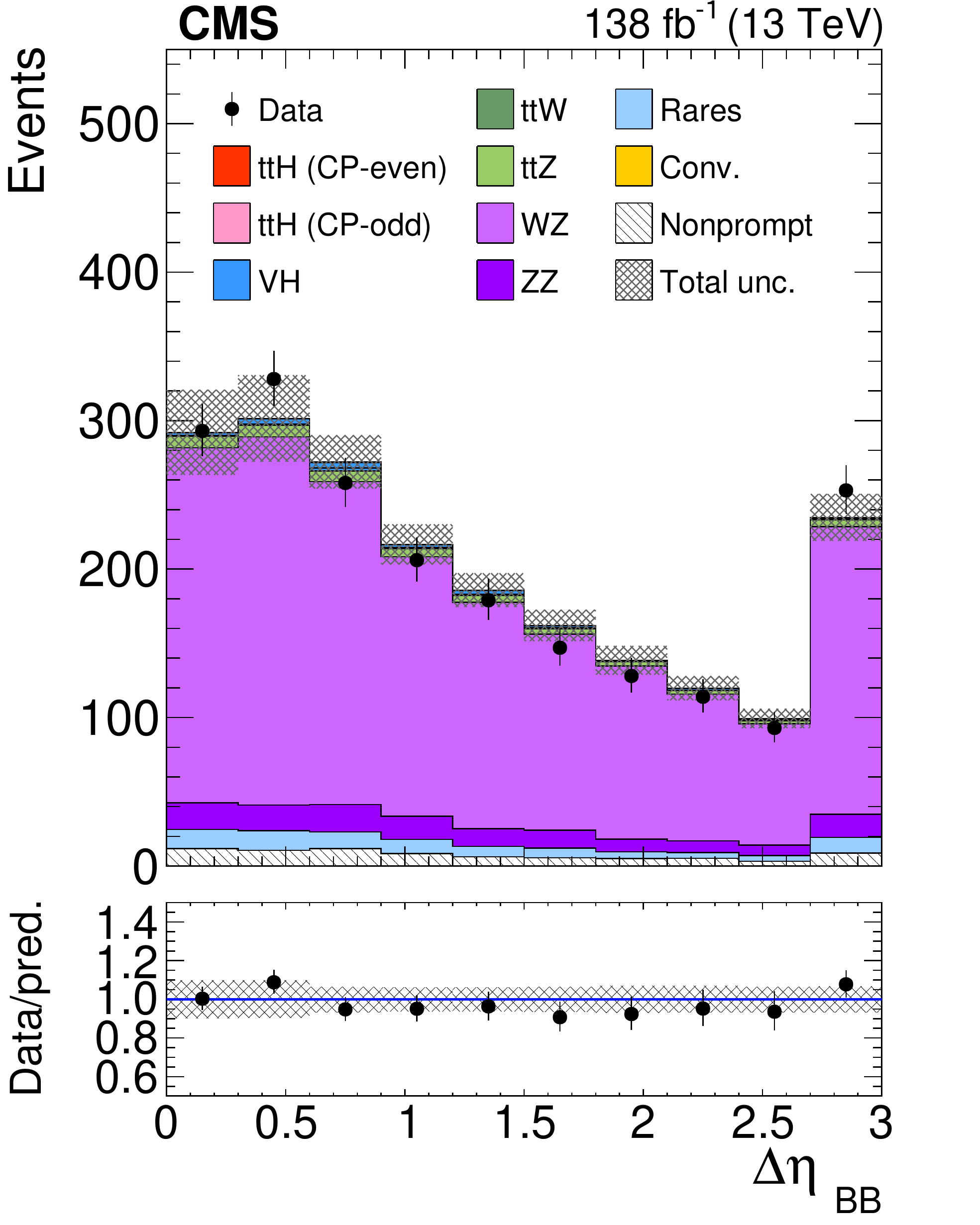}
\includegraphics*[height=60mm]{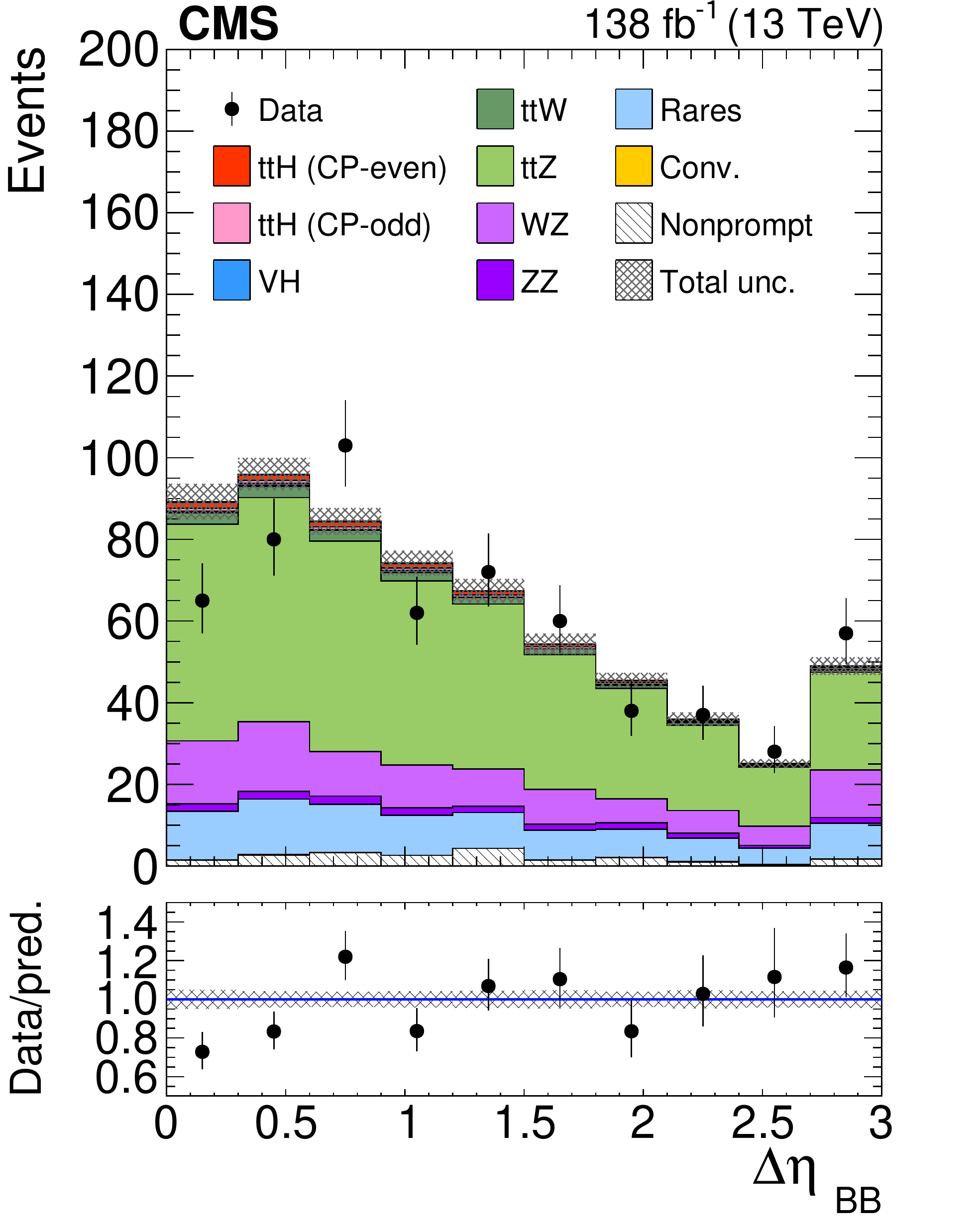}
\includegraphics*[height=60mm]{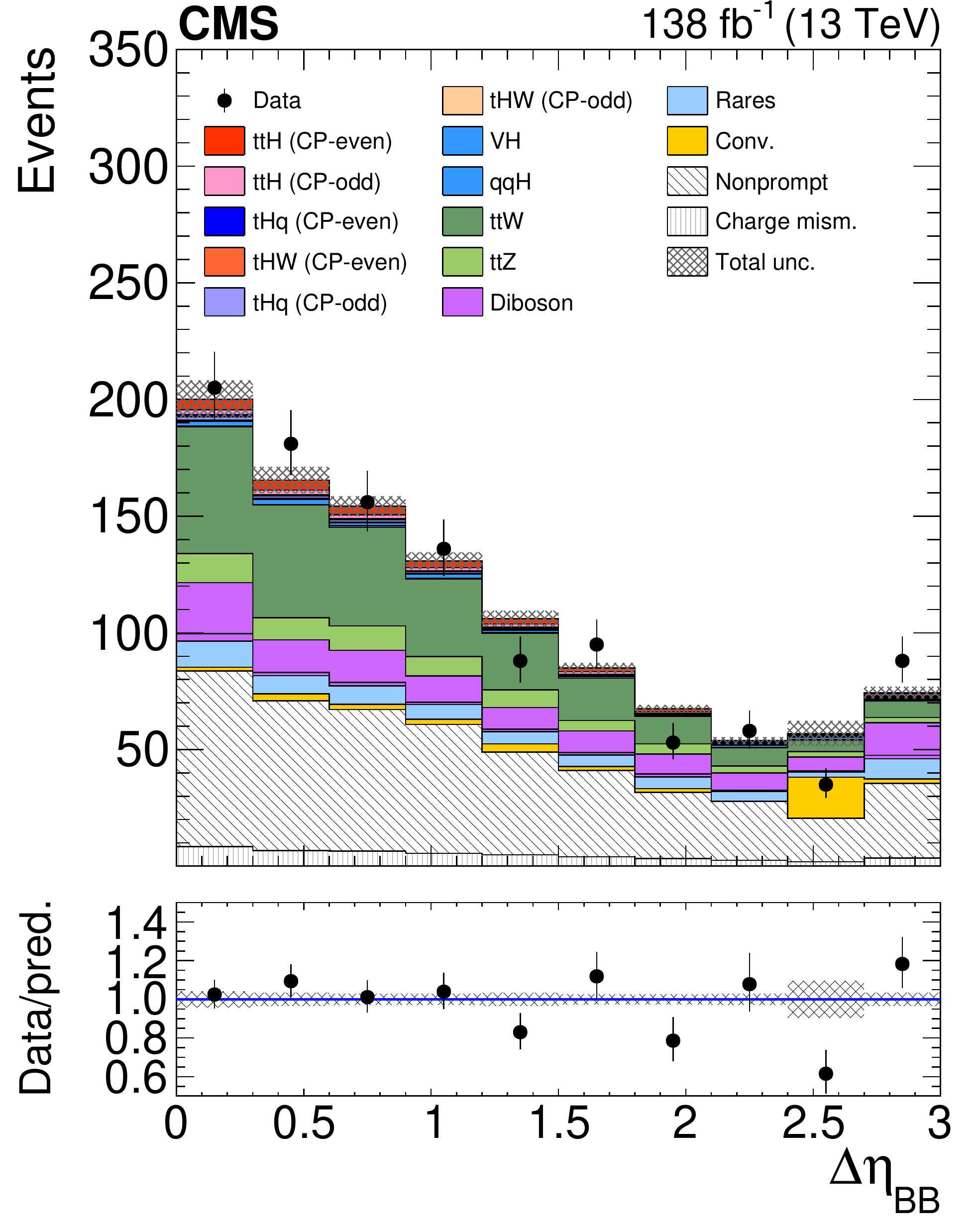}
\caption{
One of the most important input variables for the \CP discriminant, $\Delta \eta_{BB}$, in the three validation regions enriched from left to right in $\PW\PZ$, $\PQt\PAQt\PZ$ and  misidentified lepton background.  
}
\label{validationplots2}
\end{figure}

The number of input features to each BDT classifier varies between 16 to 25. Table~\ref{Table:BDTinputs} summarizes all the input features used for the BDTs. The modelling of all variables has been checked in three validation regions enriched in misidentified lepton background, $\PQt\PAQt\PZ$ and $\PW\PZ$, respectively. Some of the most important variables in the three validation regions are shown in Figures ~\ref{validationplots1} and ~\ref{validationplots2}.  

\begin{figure}
\centering
\includegraphics*[height=60mm]{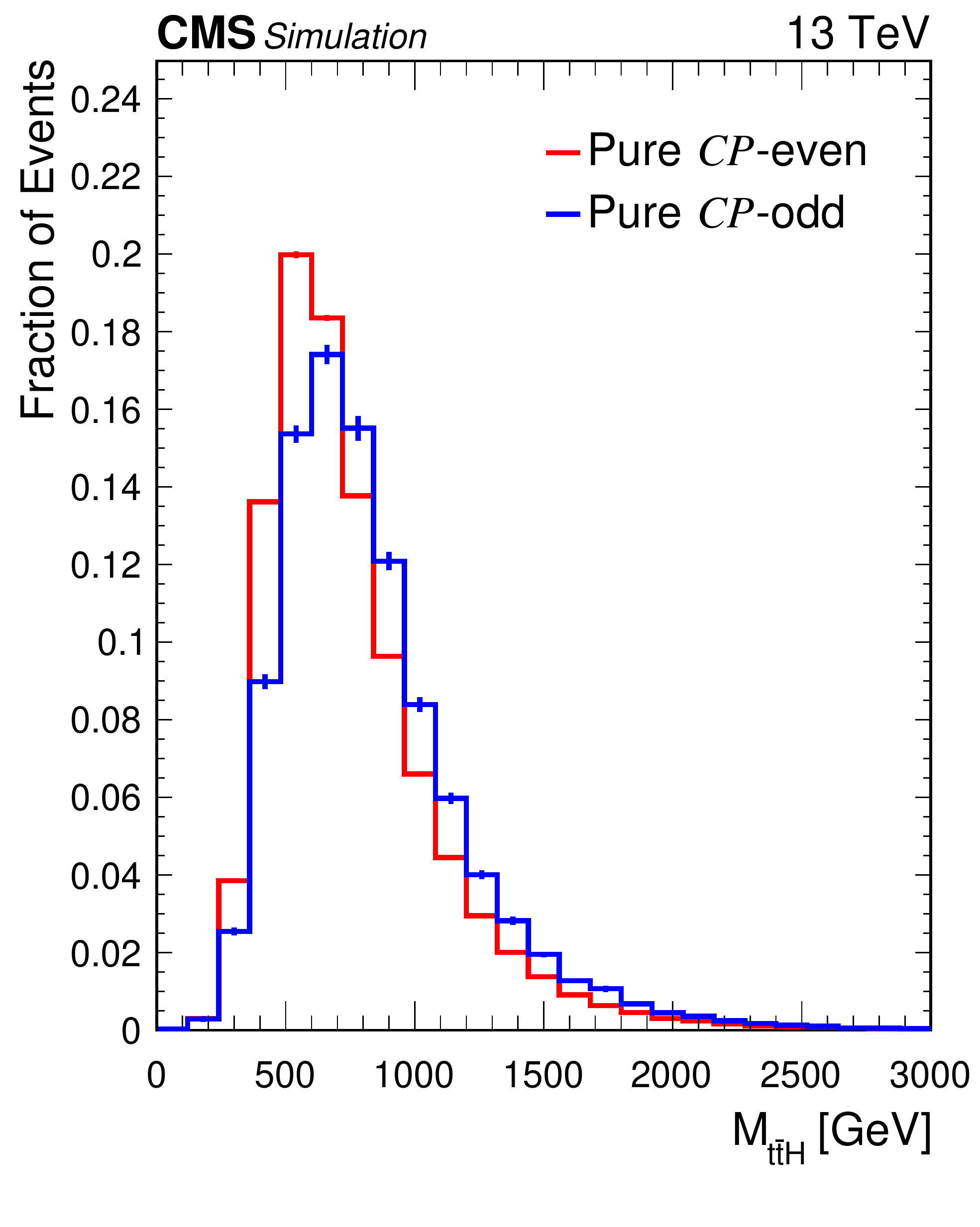}
\includegraphics*[height=60mm]{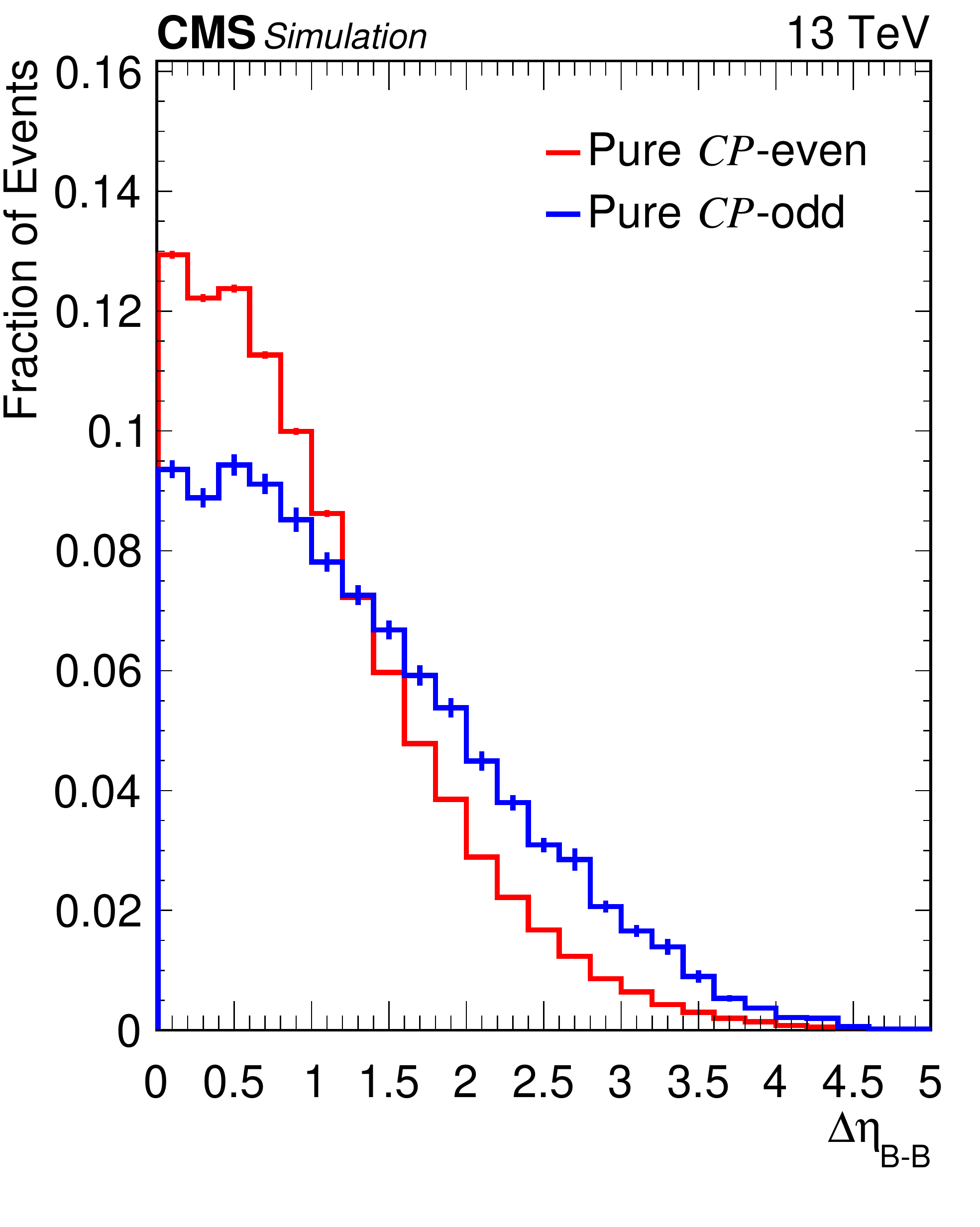}
\includegraphics*[height=60mm]{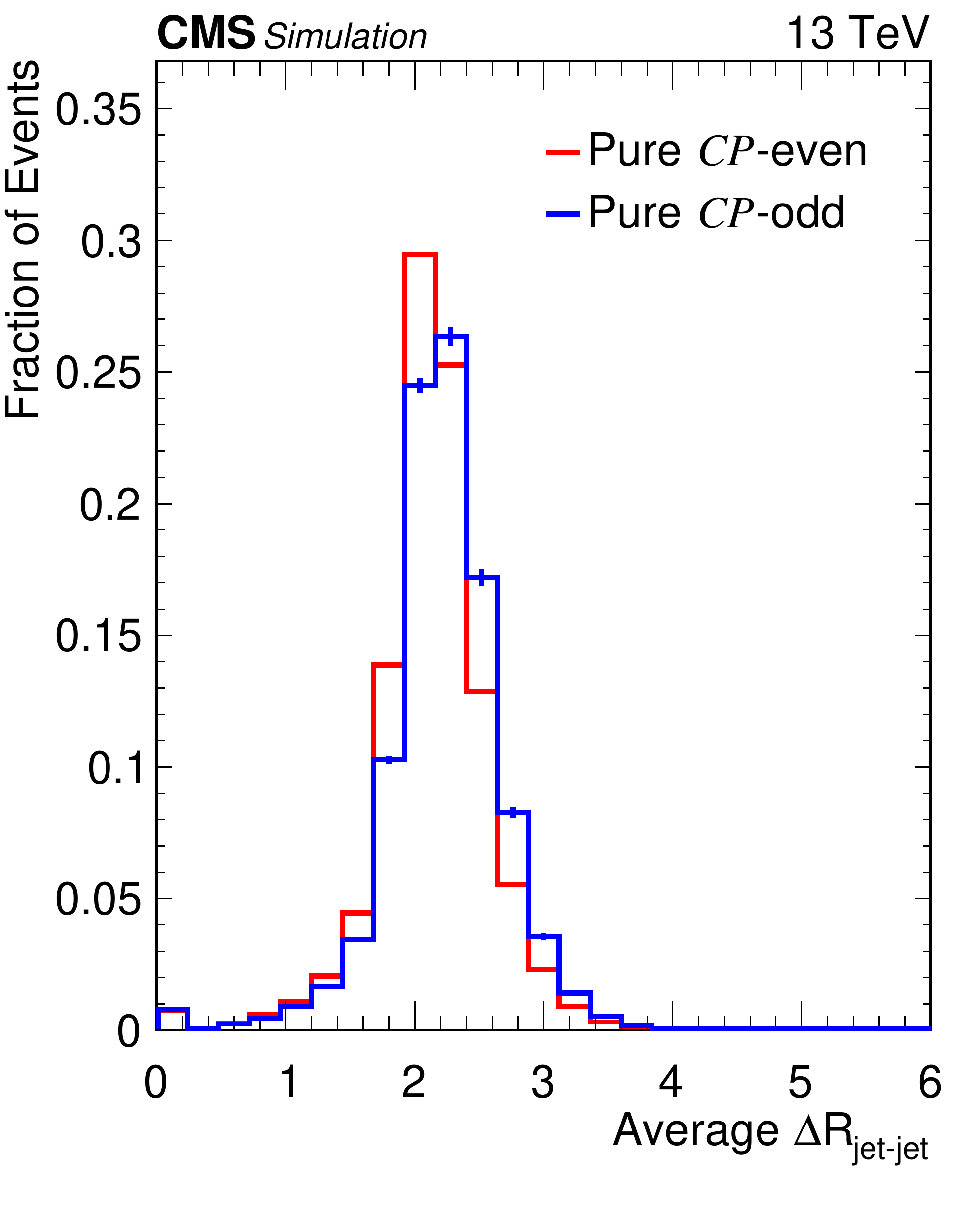}
\caption{
  Most important input variables to the \textsc{XGBoost} used for \CP discrimination in \twoLeptonssZeroTau{} channel, defined in Table~\ref{Table:BDTinputs}. The vertical bars represent the statistical uncertainty originating from the limited amount of simulated events. When not visible, the bars are smaller than the line width.
}
\label{MostImp2lss0tau}
\end{figure}

\begin{figure}
\centering
\includegraphics*[height=60mm]{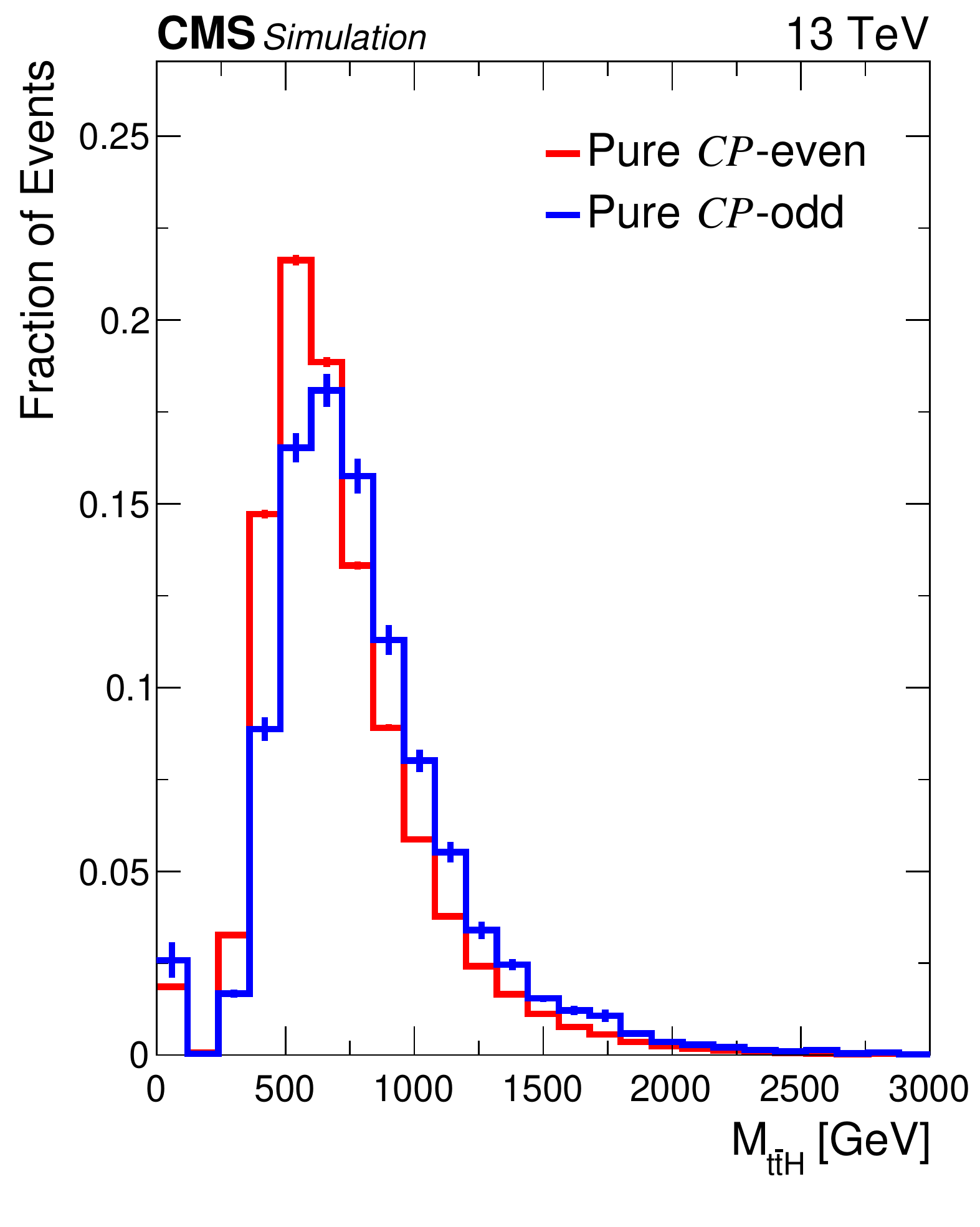}
\includegraphics*[height=60mm]{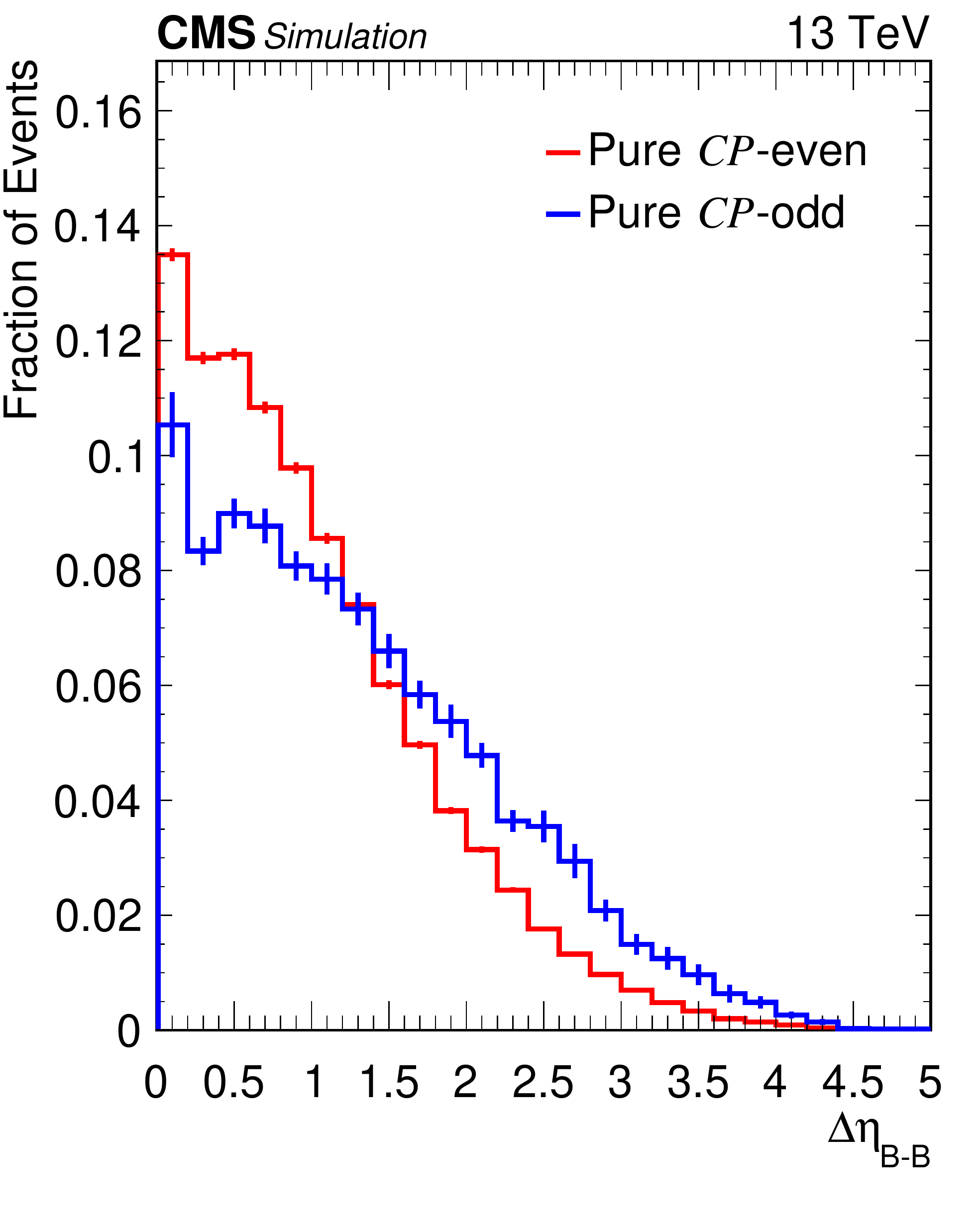}
\includegraphics*[height=60mm]{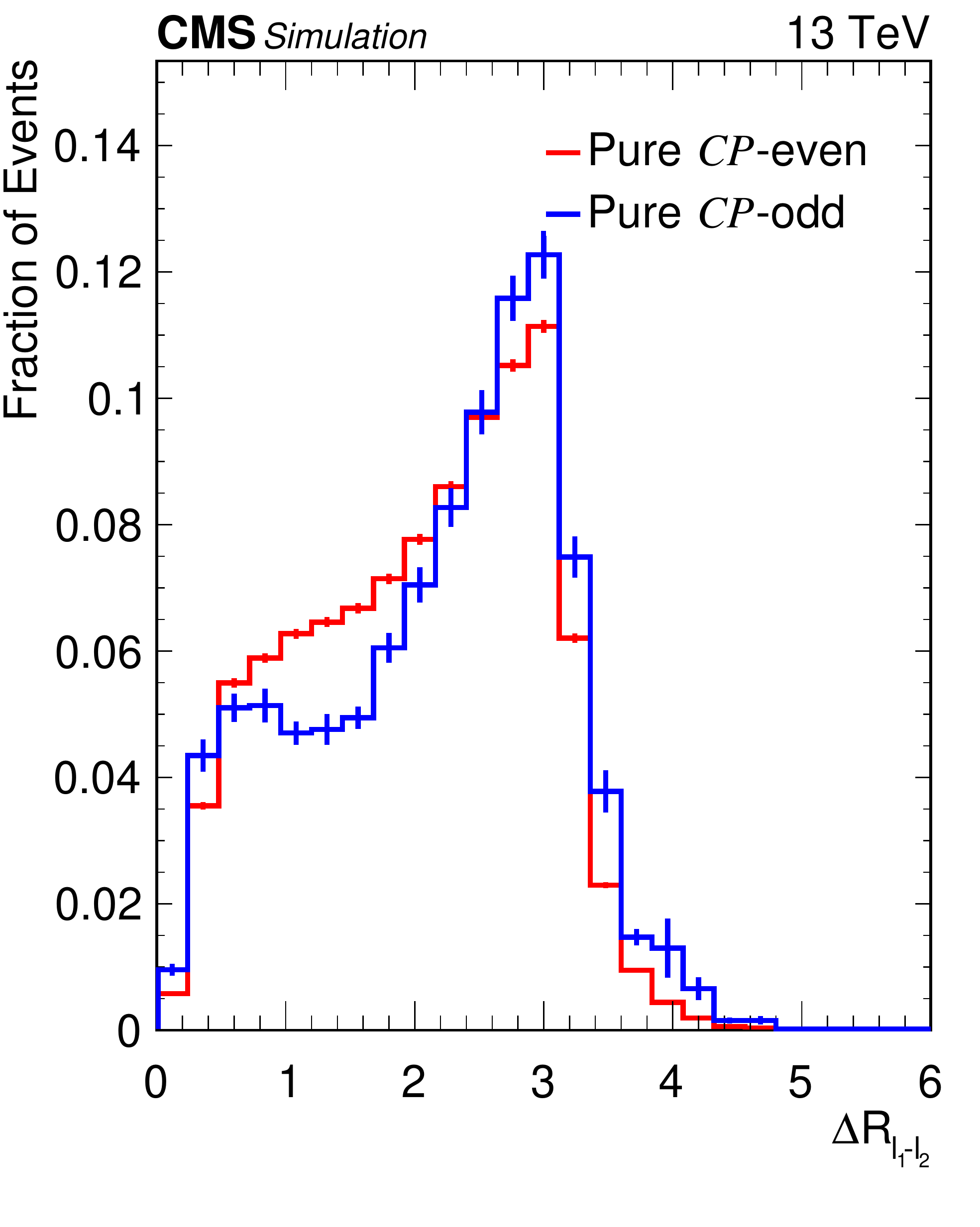}
\caption{
  Most important input variables to the \textsc{XGBoost} used for \CP discrimination in \threeLeptonZeroTau{} channel, defined in Table~\ref{Table:BDTinputs}. The vertical bars represent the statistical uncertainty originating from the limited amount of simulated events. When not visible, the bars are smaller than the line width.
}
\label{MostImp3l0tau}
\end{figure}

Some representative input feature distributions, corresponding to those with the highest post-training importance score in each final state, are shown in Figs.~\ref{MostImp2lss0tau} and~\ref{MostImp3l0tau} for the  \twoLeptonssZeroTau{} and \threeLeptonZeroTau{} channels, respectively. The variables shown are found to discriminate between the two CP scenarios, as indicated in references~\cite{Gritsan_2016} and~\cite{Demartin:2014fia}.

\begin{table}
\centering
\topcaption{Input features for the three BDTs. A check mark (\checkmark) indicates the variable is used in a given final state, whereas a long dash (\NA) indicates the variable is not used in that final state.}
\label{Table:BDTinputs}
\cmsTable{
\begin{tabular}{l c c c l}
\hline
Variable description & \twoLeptonssZeroTau{}     & \twoLeptonssOneTau{}         & \threeLeptonZeroTau{}\\
\hline  
 $\pt$ of jet 1 &  \NA                   & \NA                         & \checkmark     \\
 $\pt$ of jet 2 &  \NA                   & \NA                         & \checkmark     \\
 $\pt$ of lepton 1 &  \checkmark          & \checkmark                & \checkmark     \\
 $\pt$ of lepton 2 &  \checkmark          & \checkmark                & \checkmark     \\
 $\pt$ of lepton 3 &  \NA                   & \NA                         & \checkmark     \\
 $\pt$ of $\PGt$ lepton &  \NA                   & \checkmark                & \NA              \\
 $\eta$ of lepton 1 &  \checkmark          & \checkmark                & \NA              \\
 $\eta$ of lepton 2 &  \checkmark          & \checkmark                & \NA             \\
 $\eta$ of $\PGt$ lepton &  \NA                   & \checkmark                & \NA             \\
 $\phi$  of lepton 1 &  \checkmark          & \checkmark                & \NA             \\
 $\phi$  of lepton 2 &  \checkmark          & \checkmark                & \NA             \\
 $\phi$  of $\PGt$ lepton &  \NA                   & \checkmark                & \NA             \\
 $m_{\mathrm{T}}(l_1,\ptmiss)+ \ptmiss$ &  \checkmark          & \NA                        & \NA             \\
 $m_{\mathrm{T}}(l_2,\ptmiss)+ \ptmiss$ system &  \checkmark          & \NA                        & \NA             \\
 $\Delta R$  of lepton 1 to its closest jet & \checkmark           & \checkmark                & \checkmark     \\
 $\Delta R$  of lepton 2 to its closest jet & \checkmark           & \checkmark                & \checkmark     \\
 Invariant mass of the reconstructed \ttH{} system ($M_{\ttH{}} = \sum_i p^{\text{lep}_i}+ \vec{\pt}^{\text{miss}} +  \sum_{i \leq k} p^{\text{jet}_i*}$) & \checkmark           & \checkmark                & \checkmark     \\
 $\Delta \eta$ of two jets with highest b score in the laboratory frame ($\Delta \eta_{BB}$) & \checkmark           & \checkmark                & \checkmark     \\
 $\Delta \eta$ of the two leptons in frame of two most-likely b jets & \checkmark           & \checkmark                & \NA             \\
 $\Delta \eta$ of two jets with highest b score in the dilepton system frame & \checkmark           & \checkmark                & \NA             \\
 $\Delta \eta$ of two jets with highest b score in the $\ell_1$-$\ell_2$ system frame & \NA                   & \NA                        & \checkmark     \\
 $\Delta \eta$ of two jets with highest b score in the $\ell_1$-$\ell_3$ system frame & \NA                   & \NA                        & \checkmark     \\
 $\Delta \phi$  of the two leptons in frame of two most-likely b jets & \NA                    & \checkmark                & \NA             \\
 $\Delta \phi$  of two jets with highest b score in the dilepton system frame & \NA                    & \checkmark                & \NA             \\
 Average $\Delta R$ among all jets &  \checkmark          & \checkmark                & \NA             \\
 Jet multiplicity &  \checkmark          & \checkmark                & \NA             \\
 \ptmiss &  \checkmark          & \checkmark                & \NA             \\
 Azimuthal angle of \ptvecmiss &  \checkmark          & \checkmark                & \NA             \\
 Highest BDT score of jet triplet from \PQt&  \checkmark          & \checkmark                & \NA             \\
 Higgs jet tagger &  \NA                & \checkmark                & \NA             \\
 Angle of $\PQt\PAQt$ and \PH{} boson in \ttH{}-system & \NA     & \checkmark                & \NA             \\
 Angle between two \PQt{} in \ttbar{}-frame & \NA            & \checkmark                & \NA             \\
 $ \Delta R_{l_{3}-l_{1}} = \sqrt{(\eta_{\ell_3} - \eta_{\ell_1})^2 + (\phi_{\ell_3} - \phi_{\ell_1})^2}$& \NA                   & \NA                        & \checkmark     \\
 $ \Delta R_{l_{1}-l_{2}} = \sqrt{(\eta_{\ell_1} - \eta_{\ell_2})^2 + (\phi_{\ell_1} - \phi_{\ell_2})^2}$& \NA                   & \NA                        & \checkmark     \\
 $ \Delta R_{l_{2}-l_{3}} = \sqrt{(\eta_{\ell_2} - \eta_{\ell_3})^2 + (\phi_{\ell_2} - \phi_{\ell_3})^2}$& \NA                   & \NA                        & \checkmark     \\
 $\eta_{\mathrm{jet1}} - \eta_{\mathrm{jet2}}$ & \NA                   & \NA                        & \checkmark     \\
 $\pt^{\mathrm{jet1}} + \pt^{\mathrm{jet2}} + \pt^{\mathrm{jet3}} + \ptmiss$ & \NA                   & \NA                        & \checkmark     \\
[\cmsTabSkip]
Total number of variables & 19          & 25                        & 16             & \\
\hline
\end{tabular}
}
\footnotesize{$^{*}$ k = 6 (4) in the \twoLeptonssZeroTau{} (\twoLeptonssOneTau{} and \threeLeptonZeroTau{}) final state} \\
\end{table}

\section{Systematic uncertainties}
\label{sec:uncertainties}

Various imprecisely-known effects may alter the event yield of the \ttH{} and \tH{} signals and of background processes,
as well as the shape of the distributions in the discriminating observables that are used for the signal extraction.
In this section, these effects, usually called systematic uncertainties, are described and their impact on the yields in the various final states is discussed. 

The systematic uncertainties are modelled as nuisance parameters in a maximum likelihood (ML) fit to set limits on BSM scenarios with modified Higgs couplings using the method described in Ref.~\cite{CMS-NOTE-2011-005}.
The fit is described in detail in Section~\ref{sec:signalExtraction}.
The nuisance parameters are allowed to modify the event yield, accounting also for the migration of events among regions and among different bins in the distributions fitted in each region. 
Nuisance parameters accounting for the limited amount of simulated events are treated using the \textit{Barlow-Beeston-lite} approach~\cite{Barlow:1993dm}.

We consider two categories of nuisance parameters: those that purely affect the yield in a category (\textit{rate uncertainties}) are assigned a log-normal probability density function,
whereas those that affect also the shape of the distributions (\textit{shape uncertainties}) are modelled via a polynomial interpolation with a Gaussian constraint and are also allowed to change the
event yields in a category.
The correlations between the various experimental uncertainty sources across the three years of data taking are detailed in the text and summarized in Table~\ref{tab:syst_correlations}. Theoretical uncertainties are treated as correlated between the three years of data taking.
\subsection{Experimental uncertainties}

 The integrated luminosities for the 2016, 2017, and 2018 data-taking years have 1.2--2.5\% individual uncertainties~\cite{CMS:2021xjt,CMS-PAS-LUM-17-004,CMS-PAS-LUM-18-002}, while the overall uncertainty for the 2016--2018 period is 1.6\%. These effects are applied as rate uncertainties.

Uncertainties in the instantaneous luminosity and in the \pp{} inelastic cross section may affect the event yield of the \ttH{} and \tH{} signals and of backgrounds obtained from the MC simulation,
because different PU conditions may modify the efficiency to pass the event selection criteria.
We vary the product of instantaneous luminosity and \pp{} inelastic cross section, referred to as the number of PU interactions, by 5\%~\cite{Sirunyan:2018nqx}, and propagate the effect to the event yields,
obtaining an uncertainty in the yield smaller than 1\%.

For the 2016 and 2017 data-taking periods, additional corrections to the ECAL measurements are applied to take into account the inefficiency in the trigger caused by the gradual shift in the timing of the inputs of the ECAL level 1 (L1) trigger in the region at $\abs{\eta} > 2.0$~\cite{CMS:2020cmk}. Correction factors are computed from data and applied to the acceptance evaluated by simulation. The uncertainty in these corrections is propagated to the final discriminators and taken as a shape uncertainty, uncorrelated across data-taking periods.

The efficiencies of triggers based on the presence of two or three electrons or muons are measured as a function of the lepton multiplicity using samples of \ttbar{} and diboson events that have been recorded using triggers based on \ptmiss. The uncertainties on the trigger efficiencies result in uncertainties in the range 1--2\%, once propagated to the event yields. For the \twoLeptonssZeroTau{} and \twoLeptonssOneTau{} categories, the uncertainty is taken as shape systematics, to account for effects in \pt and $\eta$ and is correlated across categories. For the \threeLeptonZeroTau{} category, the uncertainty is taken as a flat normalization and treated as uncorrelated with the other two categories. Trigger uncertainties are treated as uncorrelated across years for all categories. 

Identification and isolation efficiencies for electrons and muons are estimated as a function of the lepton \pt and $\eta$ in a dedicated way for the different lepton selections, as detailed in Ref.~\cite{Sirunyan:2020icl},
and their uncertainties are propagated to the final results.
The total uncertainties in the muon and electron identification and isolation efficiencies are taken as shape uncertainties and are correlated across years, amounting to less than 2\%.
The uncertainty in the $\tauh$ identification efficiency amounts to about 5\%: it is dominated by statistical effects and is therefore treated as uncorrelated across years, $\pt$ bins, and decay modes.

The energy scales of electrons and muons are known with an uncertainty of less than 1\% and are neglected in the analysis.
The uncertainties associated to the energy scale of $\tauh$ are obtained by varying the corresponding scale factor by its uncertainties, which are lower than 1.1\%~\cite{Sirunyan:2018pgf}.  
The total uncertainty in the $\tauh$ energy scale is taken as a shape uncertainty and is treated as uncorrelated across years.

The uncertainty in the jet energy scale (JES) is measured to an uncertainty of a few percent and propagated to the event yields as shape uncertainties~\cite{Khachatryan:2016kdb}.
The uncertainty in the jet energy resolution (JER) is smaller than the effect of the uncertainty in the JES.
Since this analysis is moderately sensitive to the JES and JER corrections, eleven independent uncertainty components are considered, grouped by the detector region.
Some of these components are correlated across the years and some are not, while all of them are treated as uncorrelated with eachother.
  
Uncertainties in the \PQb{} tagging efficiencies and mistag rates as function of the jet $\pt$ and $\eta$ are propagated to the event yields
by considering three sources: the impact of the JES uncertainties, the purity in the control sample used to derive the data-to-MC correction factors, and the statistics of the sample used to derive the data-to-MC correction factors~\cite{Sirunyan:2017ezt}.
The effect of the uncertainties in the \PQb{} tagging efficiency and mistag rate is small compared to the statistical uncertainties.
The \PQb{} tagging efficiency uncertainties of statistical origin are uncorrelated across years, while the experimental ones are correlated.
    
Uncertainties in the $\ptmiss$ resolution and response are taken into account by varying the jet energy scale and resolution within their respective uncertainties
and recomputing \ptmiss{} and all \ptmiss{} related observables after each variation. The resulting total uncertainty is taken as a shape uncertainty and is treated as correlated across years.  An additional uncertainty in \ptmiss{} is obtained by varying the momentum of particle flow candidates according to their resolution. 

The procedures used to compute the uncertainties in the misidentified leptons background estimation are detailed in Ref.~\cite{Sirunyan:2020icl}.
These uncertainties amount to about 10--20\%, depending on the category.
An additional uncertainty is added in the \twoLeptonssOneTau{} region, because not all effects are well encapsulated by the current systematic uncertainties in the estimated \tauh misidentification probability.

The uncertainty in the normalization of the charge flip background, which is present in the \twoLeptonssZeroTau{} and \twoLeptonssOneTau{} categories, amounts to 30\%, coming mainly from electrons.

The uncertainties associated to the misidentified leptons background estimation techniques are treated as uncorrelated across the years.
  
\subsection{Theoretical uncertainties}

The signal rates are measured in units of the SM \ttH{} and \tH{} production rates:
the measurement is therefore affected by  uncertainties in the \ttH{} and \tH{} cross sections.

The uncertainty in the SM \ttH{} (\tH{}) cross section, computed at NLO accuracy, amounts to $^{+6.8\%}_{-10.0\%}$ ($^{+4.2\%}_{-6.8\%}$), of which $^{+5.8\%}_{-9.3\%}$ ($^{+4.1\%}_{-6.7\%}$) are due to missing higher orders and 3.6\% arises from uncertainties in the parton distribution function (PDF) and the strong coupling $\alpha_{s}$~\cite{Heinemeyer:2013tqa}.

The uncertainty applied on the normalization of non-SM scenarios is the same as the SM one. The \ttH{} process is simulated at LO in this analysis. An uncertainty in the acceptance due to missing higher orders treated as a shape systematic is considered to cover for discrepancies between NLO and LO simulation.

The uncertainties in the branching fraction for the \PH{} boson to decay into $\PW\PW$ ($\Pgt\Pgt$) are taken into account, amounting to 1.54 (1.65)\%~\cite{deFlorian:2016spz}.

Uncertainties in the acceptance that are due to missing higher orders are treated as shape systematics, as they affect the shape of the distribution in the observable used for signal extraction.
They are estimated by varying the renormalization ($\mu_{\textrm{R}}$) and factorization ($\mu_{\textrm{F}}$) scales  between 0.5 and 2 times their default values, with the constraint that $0.5 \leq \mu_{\textrm{F}}/\mu_{\textrm{R}} < 2$, as recommended in~\cite{Cacciari:2003fi,Catani:2003zt,Frederix:2011ss}. The effect of uncertainties in the PDFs on these distributions is evaluated as recommended in~\cite{Butterworth:2015oua}.

All uncertainties associated to theoretical calculations are considered correlated across years.

The extrapolation of the $\PW\PZ$+jets and $\PZ\PZ$+jets background rates from the 3$\ell$ and 4$\ell$-CRs to the SR
depends on the heavy-flavour content of $\PW\PZ$+jets and $\PZ\PZ$+jets background events. According to the
MC simulation, most of the \PQb{} jets reconstructed in  $\PW\PZ$+jets and $\PZ\PZ$+jets background events arise from
the misidentification of light-flavour quark and gluon jets rather than from charm or bottom quarks.
We assign an uncertainty of 40 (10)\%  on $\PW\PZ$+jets and $\PZ\PZ$+jets events with at least (less than) two \PQb-tagged jets, to cover the modelling of the heavy-flavour content in these 
background events. Besides, in order to account for the differences in the jet multiplicity distribution between
data and simulation in the 3$\ell$ CR described in Section~\ref{sec:anal_strategy}, we apply a 30\% uncertainty on $\PW\PZ$+jets and $\PZ\PZ$+jets events with more than two jets. 

The uncertainty in the background arising from production of top quark pairs in association with real or virtual photons amounts to 30\%~\cite{Sirunyan:2020icl}.
An uncertainty of 50\% is assigned to other rare backgrounds that are modelled using MC simulation~\cite{Sirunyan:2020icl}. These choices account for the extrapolation from the inclusive phase space to the phase space relevant for this analysis, in particular to events with a high multiplicity of jets and b-tagged jets, as required to pass the event selection criteria.
  
We make no assumption on the rates of the irreducible $\PW\PZ$+jets, $\PZ\PZ$+jets, $\PQt\PAQt\PW$, $\PQt\PAQt\PW\PW$, and $\PQt\PAQt\PZ$ backgrounds:
the normalization of these backgrounds is determined simultaneously in the ML fit used for the signal extraction, described in Section~\ref{sec:signalExtraction}. The total normalization for each process is constrained by the dedicated CRs defined in Section~\ref{sec:anal_strategy}. 

The largest impacts on the measurement of the coupling modifiers are due to: the statistical uncertainty of observed data; the uncertainties related to the estimation of the misidentified leptons and flips backgrounds; the uncertainties on the jet reconstruction; the theoretical uncertainties, which affect the yield and the distribution of the discriminating observables for the ttH and tH signals as well as for the main irreducible backgrounds, arising from $\PQt\PAQt\PW$, $\PQt\PAQt\PW\PW$, and $\PQt\PAQt\PZ$, and $\PQt\PZ$ production; and the uncertainty applied to cover for modelling differences between LO and NLO simulation of the signal.

\begin{table}
\centering
\topcaption{
  Summary of the uncertainty sources, their type, and their correlations across the three data-taking periods.  Trigger efficiency uncertainty is taken as a shape or normalization systematic depending on the channel. 
}
\label{tab:syst_correlations}
\cmsTable{
\begin{tabular}{lcc}
\hline
Uncertainty source & Type & Correlation 2016--2018 \\[\cmsTabSkip]
\multicolumn{2}{l}{Experimental uncertainties} & \\
Trigger efficiency 									& Norm. / Shape 		& Uncorrelated 						\\
Integrated luminosity										& Norm. 				& Uncorr. / Corr.		 			\\
\PQb{} tagging efficiency and mistag rate 				& Shape 				& Uncorr. (stat.) / Corr. (exp.) 			\\
Pileup reweighting                                          &        Shape                                       &   Correlated    \\ 
L1 ECAL trigger inefficiency 								& Shape 				& Uncorrelated						\\
Jet energy scale and resolution 									& Shape 				& Uncorr. / Corr.					\\
Energy scale of $\tauh$ 				& Shape 				& Uncorrelated						\\
Identification and isolation efficiency for $\Pe$  	& Shape 				& Correlated 						\\
Identification and isolation efficiency for $\Pgm$ 	& Shape 				& Correlated 						\\
Identification efficiency for $\tauh$ 						& Shape 				& Uncorrelated						\\
Misidentified lepton background rate								& Shape				& Correlated 						\\
Flip background rate									& Norm.				& Correlated 						\\
$\ptmiss$ resolution and response 						& Shape 				& Correlated						\\[\cmsTabSkip]
\multicolumn{2}{l}{Theoretical uncertainties} & \\
$\Pgm_r$/$\Pgm_f$ scale                                                  & Shape                            & Correlated \\
PDF                                                  & Norm.                            & Correlated \\
$\ttH{}$ LO vs. NLO diferences                                  & Norm.                            & Correlated \\
Branching Ratio                                   & Norm.                            & Correlated \\
Extrapolation of the $\PW\PZ$+jets and $\PZ\PZ$+jets                                   & Norm.                            & Correlated \\
$\ttbar\gamma$ Normalization & Norm.                            & Correlated \\
Rare background Normalization  & Norm.                            & Correlated \\[\cmsTabSkip]

\end{tabular}
}

\end{table}

\section{Signal extraction}
\label{sec:signalExtraction}
We perform a ML fit of the distributions of selected observables in the SRs defined for the \twoLeptonssZeroTau{}, \twoLeptonssOneTau{}, and \threeLeptonZeroTau{} categories as well as the CRs on the \ensuremath{3\ell} and \ensuremath{4\ell} categories. The regions are further separated into subcategories that target different processes by means of DNN-based classifiers, as described in Section~\ref{sec:anal_strategy}.
In the \twoLeptonssZeroTau{} category, events are further classified depending on the flavour of the leptons ($\Pe\Pe$/$\Pe\Pgm$/$\Pgm\Pgm$), except in the \ttH{} node, where no flavour categorization is applied.
In the \threeLeptonZeroTau{} category, the \tH{} and \ttH{} nodes are separated according to the number of \PQb-tagged jets.
In the node corresponding to the background-dominated region, events are also separated according to the lepton flavours ($\Pe\Pe\Pe$, $\Pe\Pe\Pgm$, $\Pe\Pgm\Pgm$, and $\Pgm\Pgm\Pgm$). 

In the three SRs (\twoLeptonssZeroTau{}, \threeLeptonZeroTau{}, and \twoLeptonssOneTau{}), the \ttH{}-like categories described above are further split depending on the score of the \CP classifiers described in Section~\ref{sec:cp_discrimination}. 

\begin{figure*}[h!]
  \centering\includegraphics[width=\textwidth]{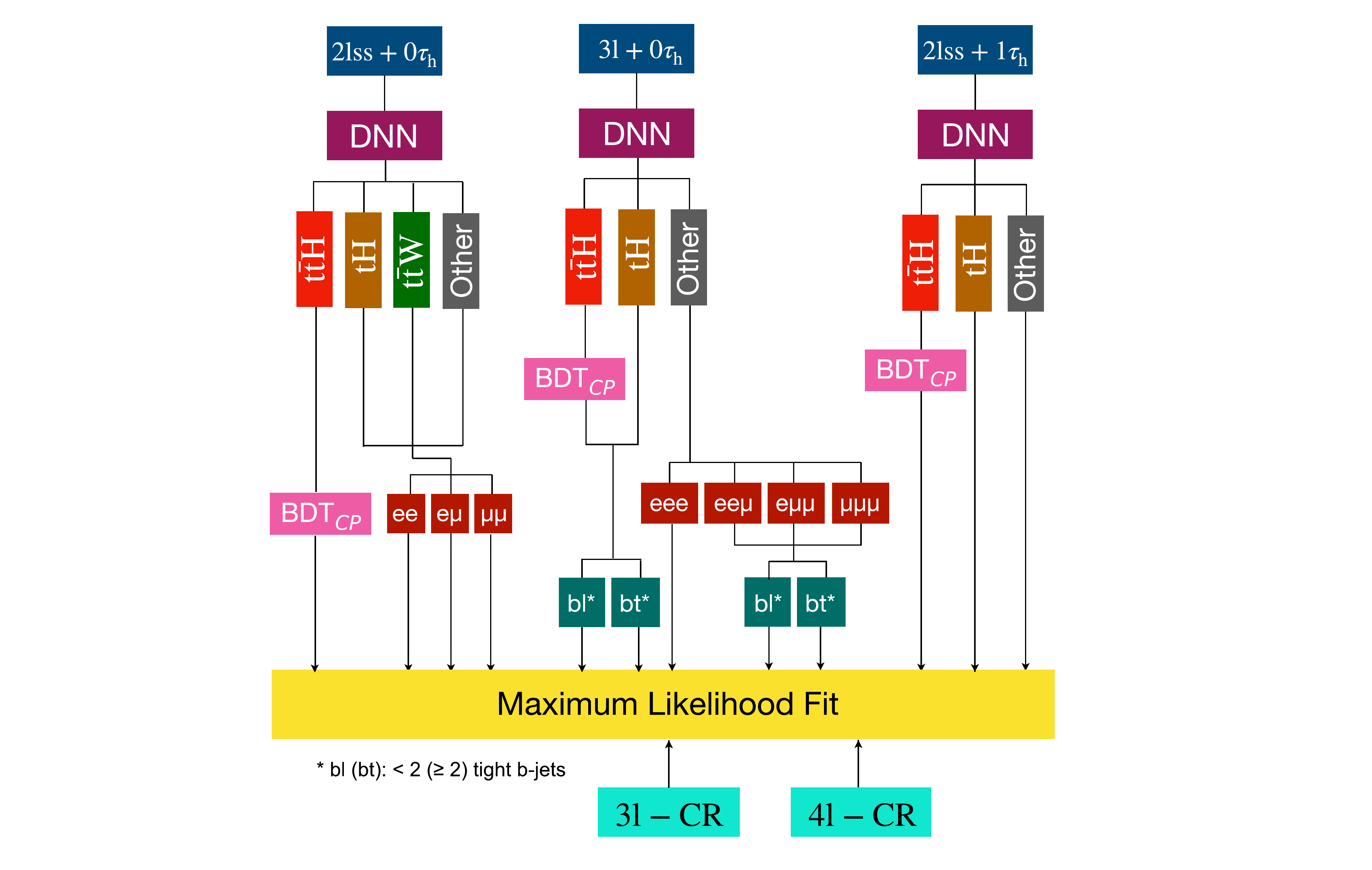}
  \caption{
    Diagram showing the categorization strategy used for the signal extraction, making use of
    MVA-based algorithms and topological variables. In addition to the three signal regions (SRs), the ML fit receives input
    from two control regions (CRs).
  }
  \label{fig:diagram}
\end{figure*}

The diagram depicted in Fig.~\ref{fig:diagram} shows the classification described above.

This categorization allowed us to exploit the different kinematic characteristics of the \ttH{} process depending on the \CP scenario. On the other hand, the dedicated \tH{} node in each of the categories allows us to constrain the normalization of this process which provides sensitivity to the sign of $\ckt$. 

\begin{figure}
  \centering
  \includegraphics[width=0.72\textwidth]{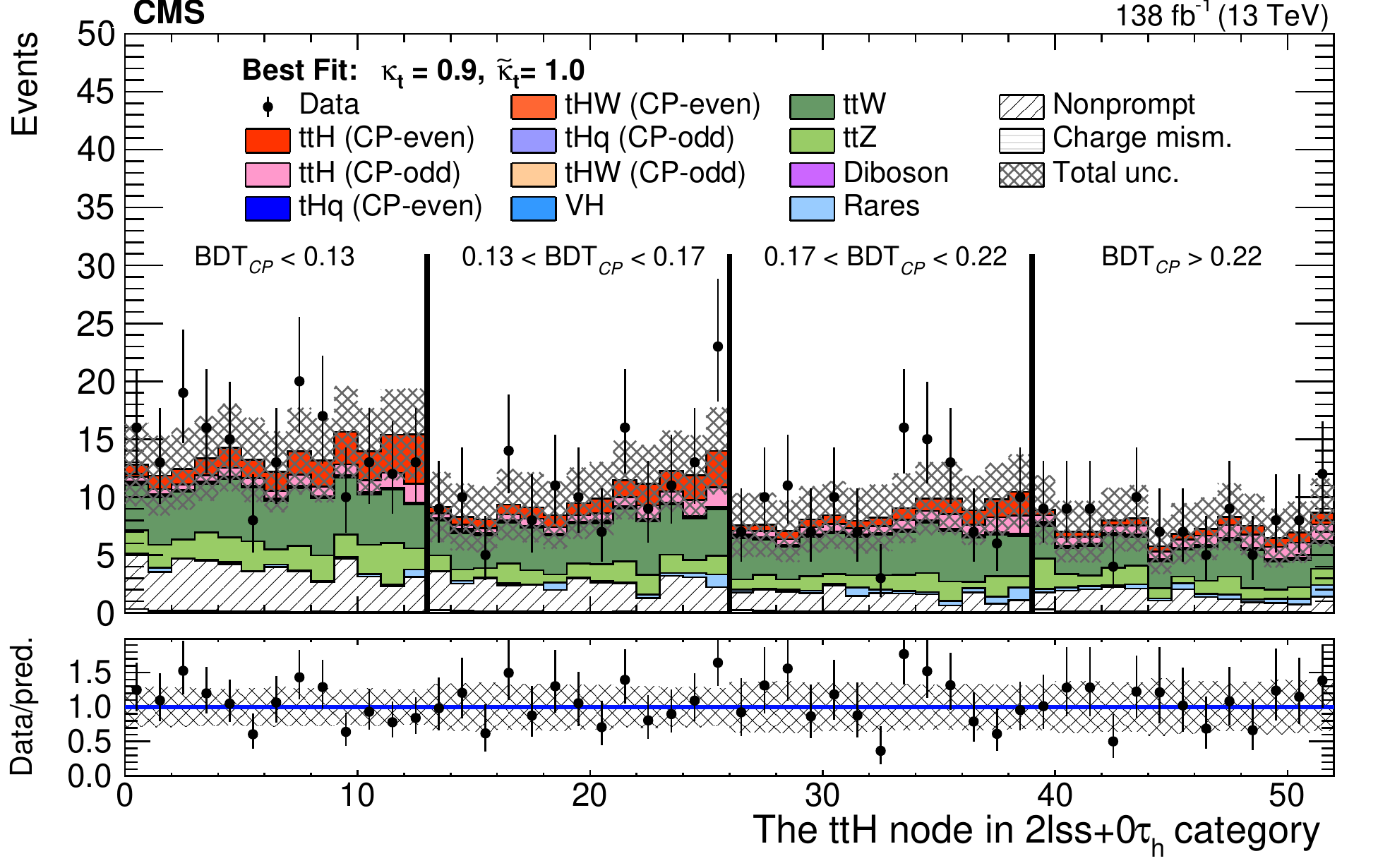}
  \includegraphics[width=0.72\textwidth]{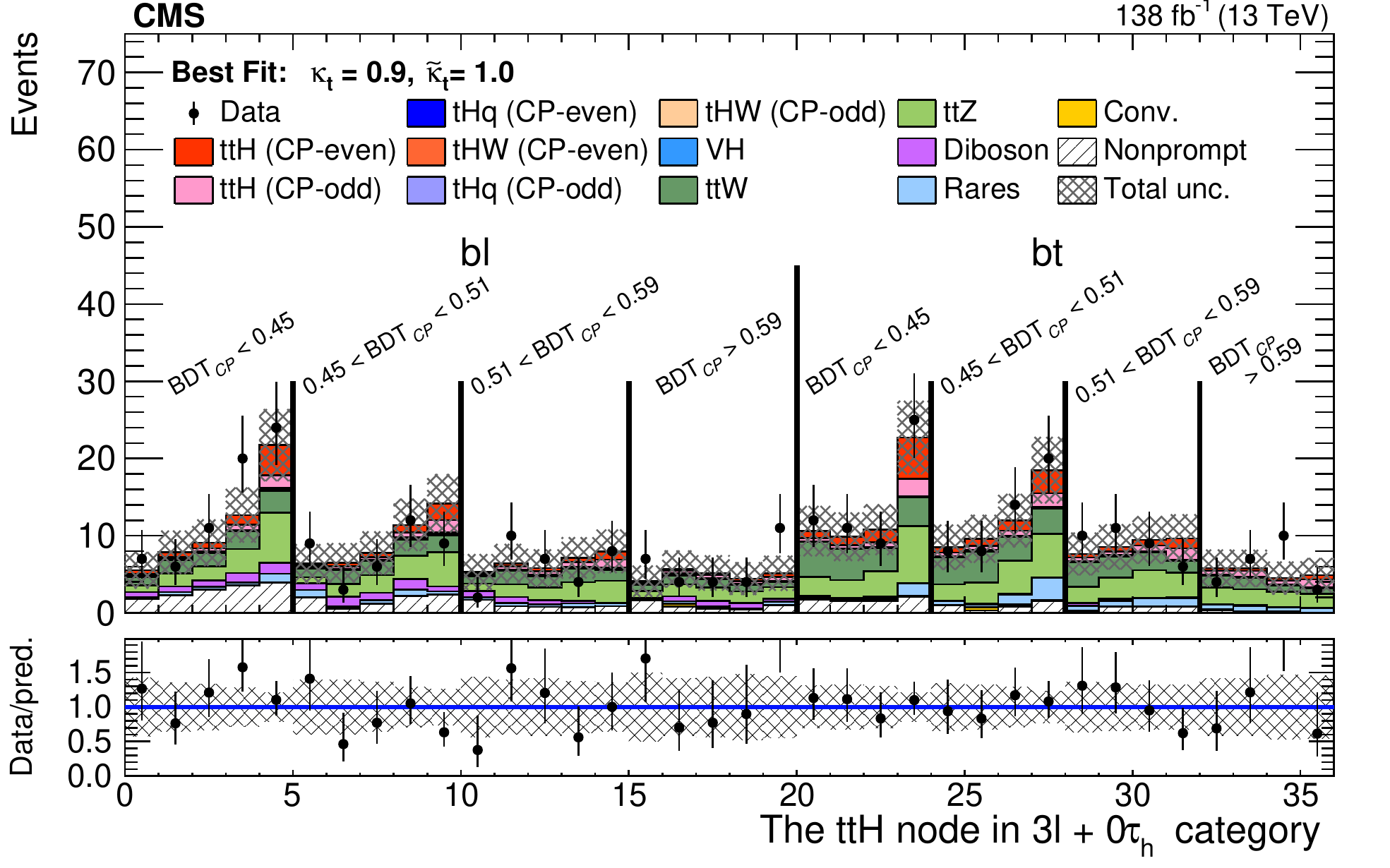}
  \includegraphics[width=0.72\textwidth]{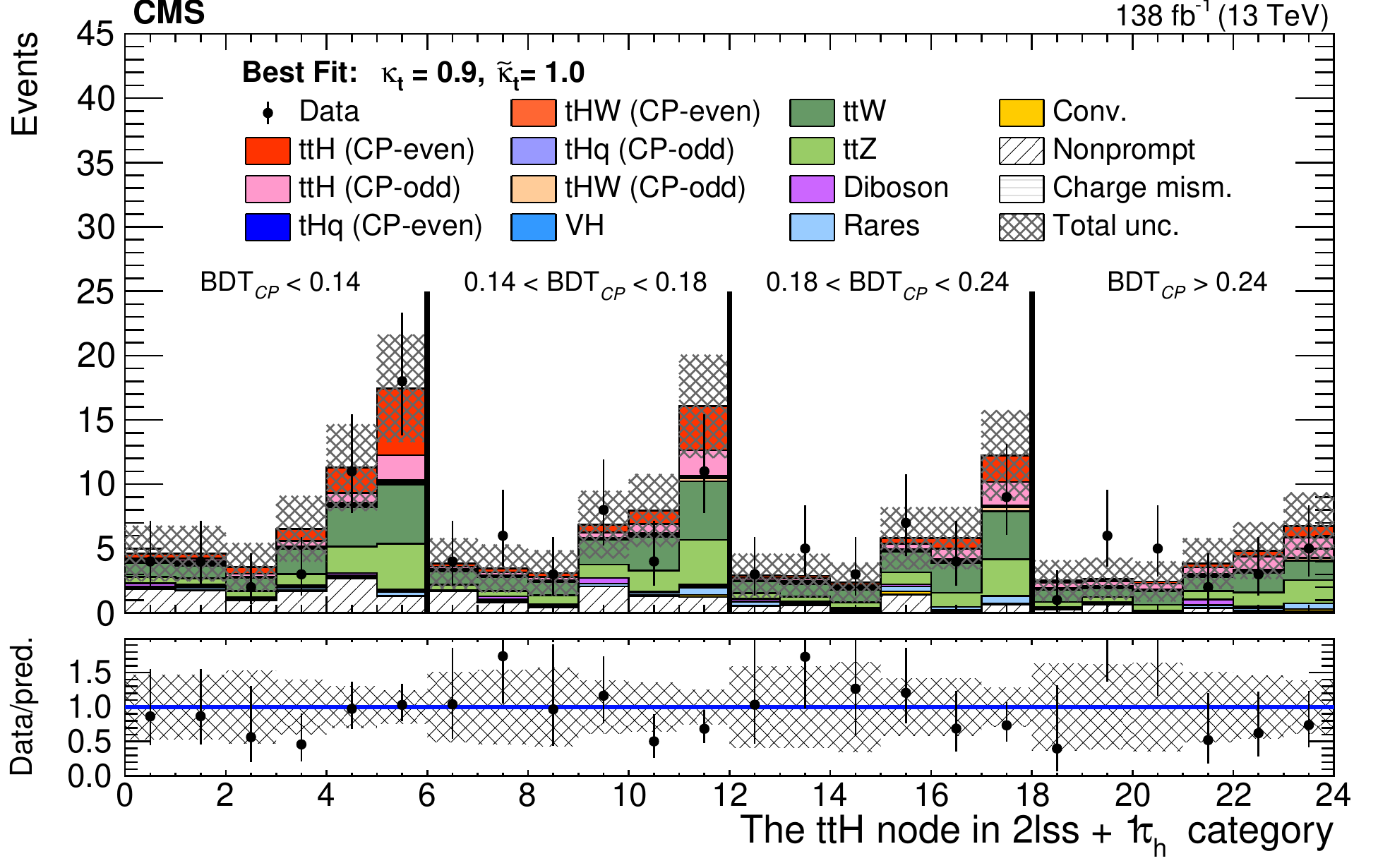}\\
  \caption{Postfit discriminating distributions used as input to the fit. Events in the \ttH{} node are categorized as described in Section~\ref{sec:signalExtraction} for the three categories:  \twoLeptonssZeroTau{} (top) \threeLeptonZeroTau{} (center) and \twoLeptonssOneTau{} (bottom). For the \twoLeptonssOneTau{} \textit{bl} (\textit{bt}) denotes events with less than (at least) two \PQb-tagged jets. The \ttH{} \CP-even (red) and \CP-odd (pink) contributions are determined from the fit. The contribution labeled as \textit{Nonprompt} refers to the backgrounds arising from misidentified leptons while the label \textit{Charge mism.}  alludes to to the backgrounds arising from  lepton charge mismeasurement. }
  \label{fig:sr_plot}
\end{figure}

The discriminating observables optimized for \ttH{} signal in order to perform the fit are shown in Fig.~\ref{fig:sr_plot}, showing the agreement between data and our statistical model evaluated at the best-fit point.  

\subsection{The BSM interpretation}

In order to explore BSM scenarios, in particular possible anomalous effects in the \PH{} boson couplings to fermions, we use the prescription introduced in Section~\ref{sec:pheno} in order to parameterize the amplitude as a function of \ckt{} and \ktt{}. These parameters modify the coupling and therefore the kinematics of \tH{} and \ttH{} processes and their normalization.
Moreover, \tH{} is sensitive to interferences of the coupling of the top quark to the \PW{} boson, in this signal extraction procedure we keep \kv{} fixed to its SM value. Branching fractions of the \PH{} boson are also set to their SM values.

\subsection{Results}

\begin{figure}
    \centering
    \includegraphics[width=0.49\textwidth]{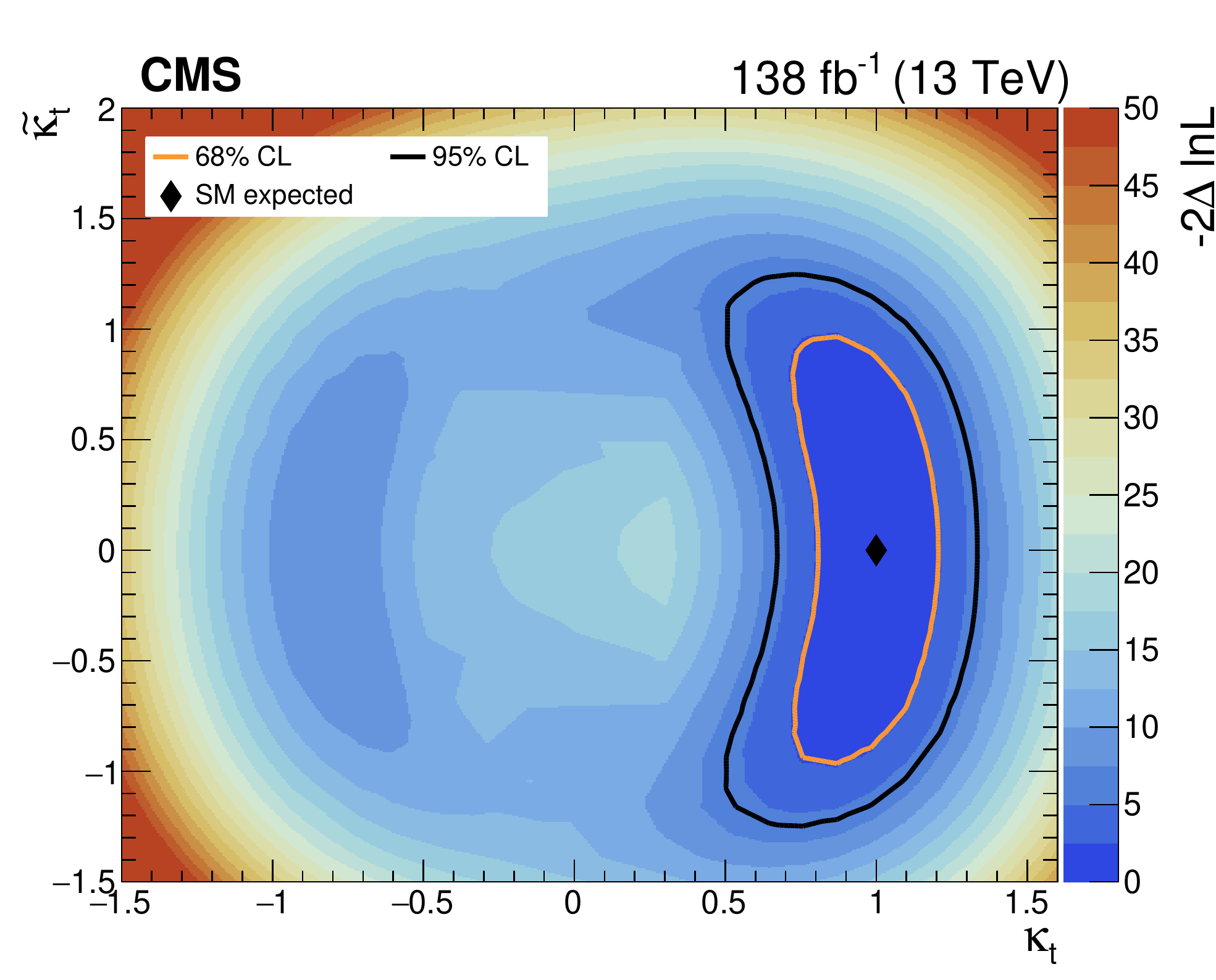}
    \includegraphics[width=0.49\textwidth]{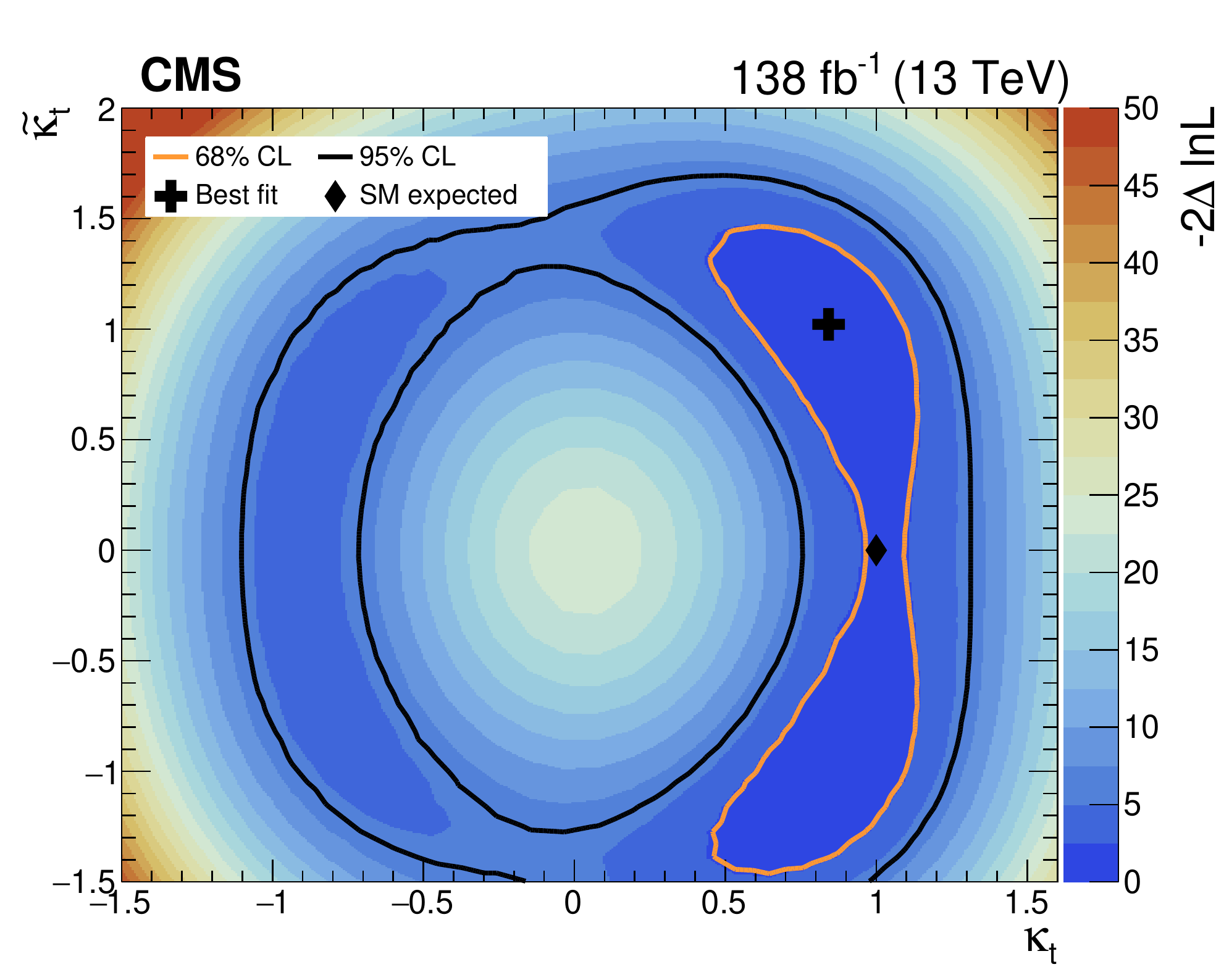}
    \caption{Likelihood scan as a function of $\ckt$ and $\ktt$: expected limits (left) and observed limits (right). The black cross shows the best value for \ckt{} and \ktt{} given by the fit. The black diamond shows the expected SM values for \ckt{} and \ktt{}. Both 68 and 95\% \CL limits are shown. \kv{} and \PH{} boson branching fractions are kept to their SM values.  }
    \label{fig:scan}
\end{figure}

Confidence regions at 68 and 95\% \CL for the $\ckt$  and $\ktt$ couplings are obtained using the strategy explained in this section, applying the asymptotic properties of the likelihood ratio.
Figure~\ref{fig:scan} (left) shows the expected likelihood scan as a function of $\ckt$ and $\ktt$. The observed likelihood scan as a function of $\ckt$ and $\ktt$ is shown in Fig.~\ref{fig:scan} (right):
the best fit is compatible with the SM within 68\% \CL. The non-zero postfit value of $\ktt$ is driven by a small excess in the most CP-odd enriched regions of the \twoLeptonssZeroTau{} and \threeLeptonZeroTau{} categories. 
 
One-dimensional confidence intervals at 68 and 95\% \CL are also obtained for \ckt{} (fixing \ktt{} to its SM value) and for \ktt (fixing \ckt{} to its SM value), as illustrated in Table~\ref{tab:CL_1d}.

\begin{table}
\centering
\topcaption{One-dimensional confidence intervals at 68 and 95\% \CL for \ckt{} (fixing \ktt{} to the SM) and  \ktt{} (fixing \ckt{} to the SM). The upper part of the table shows the expected limits while the lower part shows the observed limits.}\label{tab:CL_1d}
\begin{tabular}{lcc}
\hline
Parameter & 68\% \CL & 95\% \CL \\
\hline
& \multicolumn{2}{c}{Expected}\\
[\cmsTabSkip]
\ckt{} & (0.87, 1.14)  & (0.74, 1.27)\\
\ktt   & (-0.71, 0.71) & (-1.01, 1.01)\\
[\cmsTabSkip]
& \multicolumn{2}{c}{Observed}  \\
[\cmsTabSkip]
\ckt{} & (0.89, 1.17)   & (-1.09, -0.74) or (0.77, 1.3)\\
\ktt   & (0.37, 1.16) or (-1.16, -0.37) &  (-1.4, 1.4)\\
\hline
\end{tabular}

\end{table}

\begin{figure}
    \centering
    \includegraphics[width=0.6\textwidth]{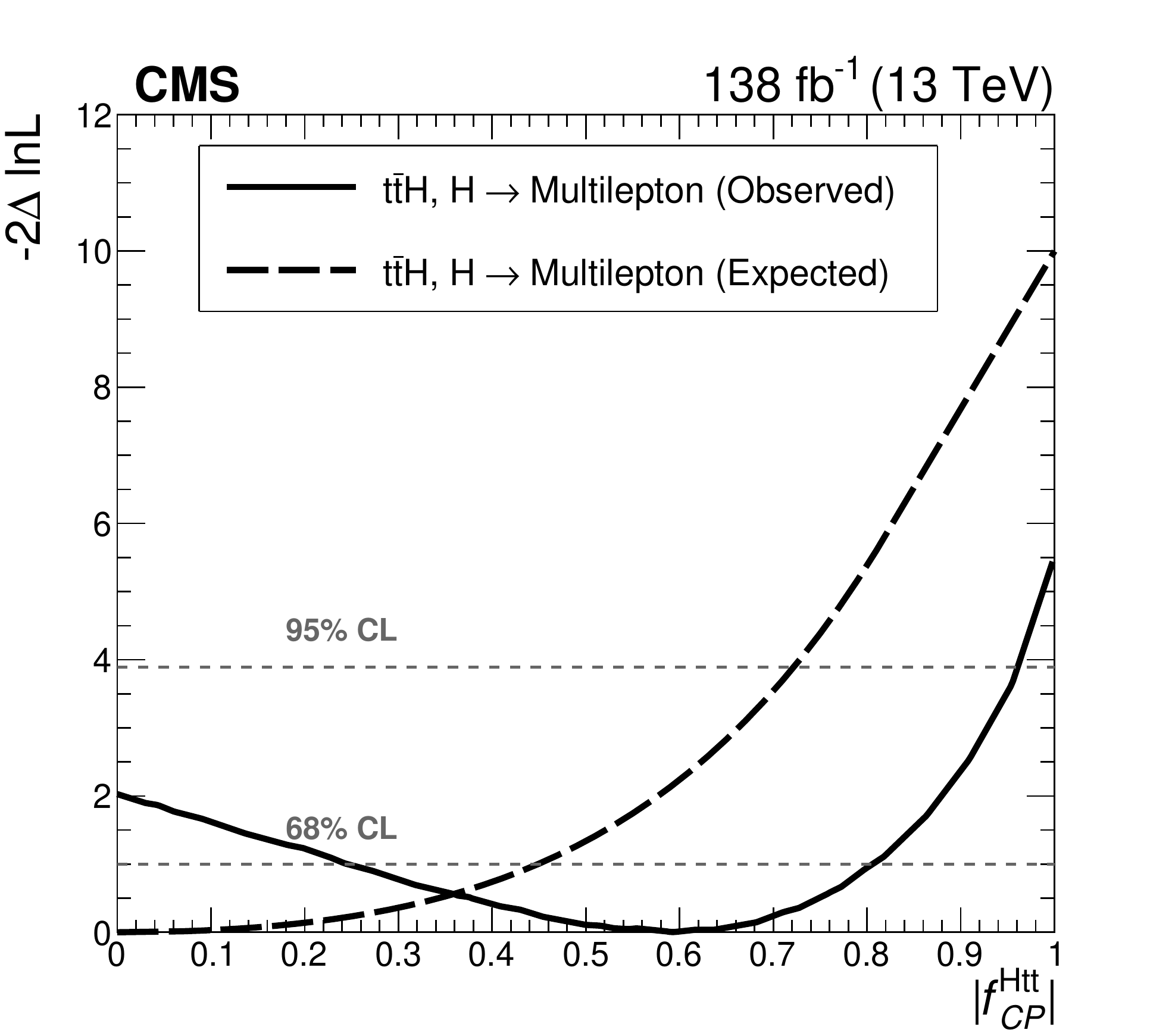}
    \caption{Likelihood scan as a function of $\fHtt$ for multilepton final estates. The solid (dashed) line shows the observed (expected) scan. }
    \label{fig:fcp_ml}
\end{figure}

We also parameterize the \ttH{} process with the signal strength multiplier to the SM \ttH{} cross section, $\mu_{\ttH}$, and a parameter $\fHtt = \frac{\abs{\ktt^2}}{(\abs{\ktt}^2+\abs{\ckt{}}^2)}$, where the \PH{} boson couplings to other particles are constrained to their SM prediction. \tH{} process is scaled with $\fHtt$ and $\mu_{\ttH}$ in the fit.
This parameterization results in a probe for a possible fractional \CP-odd contribution, yielding a best fit value of $\fHtt = 0.59$ with an interval of (0.24, 0.81) at 68\% \CL using multilepton final states only. The result is compatible with the SM \CP-even scenario at the 95\% CL. The pure \CP-odd scenario is excluded with more than two standard deviations.

\section{Combined results for \CP in ttH events}

We combine the results described in Section~\ref{sec:signalExtraction} for the multilepton decay channel with previously published CMS results on the \CP parameters in other \ttH{} decay channels, namely the $\PH\to\PGg\PGg$~\cite{Sirunyan:2020sum} and $\PH\to\PZ\PZ$ ones~\cite{Sirunyan:2021fpv}.

Uncertainties that are common between the various channels are taken as fully correlated.  We detail in Section~\ref{sec:comb:np} the treatment for all systematic uncertainties.

\subsection{Systematic uncertainties for the combination of \CP measurements}
\label{sec:comb:np}

As outlined in Section~\ref{sec:uncertainties} for the multilepton final state, we model systematic uncertainties as nuisance parameters in a ML fit to extract the  \ttH{} and \tH{} cross section and set limits on BSM scenarios with modified Higgs couplings.
The statistical model is described in detail in Section~\ref{sec:signalExtraction}.

The correlations between the various uncertainty sources across the three years of data taking are detailed in Section~\ref{sec:uncertainties} and summarized in Table~\ref{tab:syst_correlations}.

The correlations between the various uncertainty sources across the different channels that we combine, as well as the changes in the modelling of some nuisance parameters to accommodate a full combination, proceed as follows:

\begin{itemize}
\item Each of the nuisance parameters describing the uncertainty in the integrated luminosity is taken as correlated across decay channels;
\item The uncertainties caused by PU effects are taken as correlated across decay channels;
\item Uncertainties in the determination of the trigger efficiencies are taken as uncorrelated across decay channels, because the uncertainties come from independent measurements;
\item Uncertainties in the identification and isolation for electrons and muons are taken as uncorrelated across the channels. The dominant component in the multilepton channel is due to the closure of the method, as scale factors are obtained in a DY sample and we need to extrapolate them to the analysis phase space. Hence, the main contribution to the lepton uncertainty comes from a different source in the multilepton channel than in other channels;
 uncertainties in the identification and energy scale of $\tauh$s are specific to the multilepton channel;
\item The uncertainties in the JES, JER, and \PQb{} tagging efficiency and mistag rate are divided into several uncorrelated components. These uncertainties are taken as correlated across the analyses, except for the JES and JER uncertainties, where they are split in a different number of components and are therefore treated as uncorrelated;

\item Uncertainties in the $\ptmiss$ resolution and response are taken as correlated across decay channels;
\item Uncertainties in the estimation of the background from misidentified leptons and in the normalization of the charge-flip background in the multilepton final state are specific to this channel;
\item The theoretical uncertainties in the signal rates, as well as in the rates of any background that is estimated from simulation in all decay channels, are taken as correlated across decay channels;  
\item The uncertainties in the branching fractions for the \PH{} boson are taken as uncorrelated per channel and correlated per decay modes of the \PH{} boson;
\item Uncertainties in the acceptance that are due to missing higher orders are treated for each process determined from simulation as fully correlated across decay channels;
\item Uncertainties in the rate of backgrounds that were determined from analysis-specific CRs are taken as uncorrelated. This includes the contribution of the $\PW\PZ$+jets, $\PZ\PZ$+jets, and $\ttbar\gamma$ backgrounds in the multilepton channel;
\item The normalization of the irreducible $\PQt\PAQt\PW$, $\PQt\PAQt\PW\PW$, and $\PQt\PAQt\PZ$ backgrounds is determined simultaneously in the ML fit used for the signal extraction, described in Section~\ref{sec:signalExtraction}.

\end{itemize}   

\subsection{Results for the combination of \texorpdfstring{\CP}{CP} parameters}

\begin{figure}
    \centering
    \includegraphics[width=0.45\textwidth]{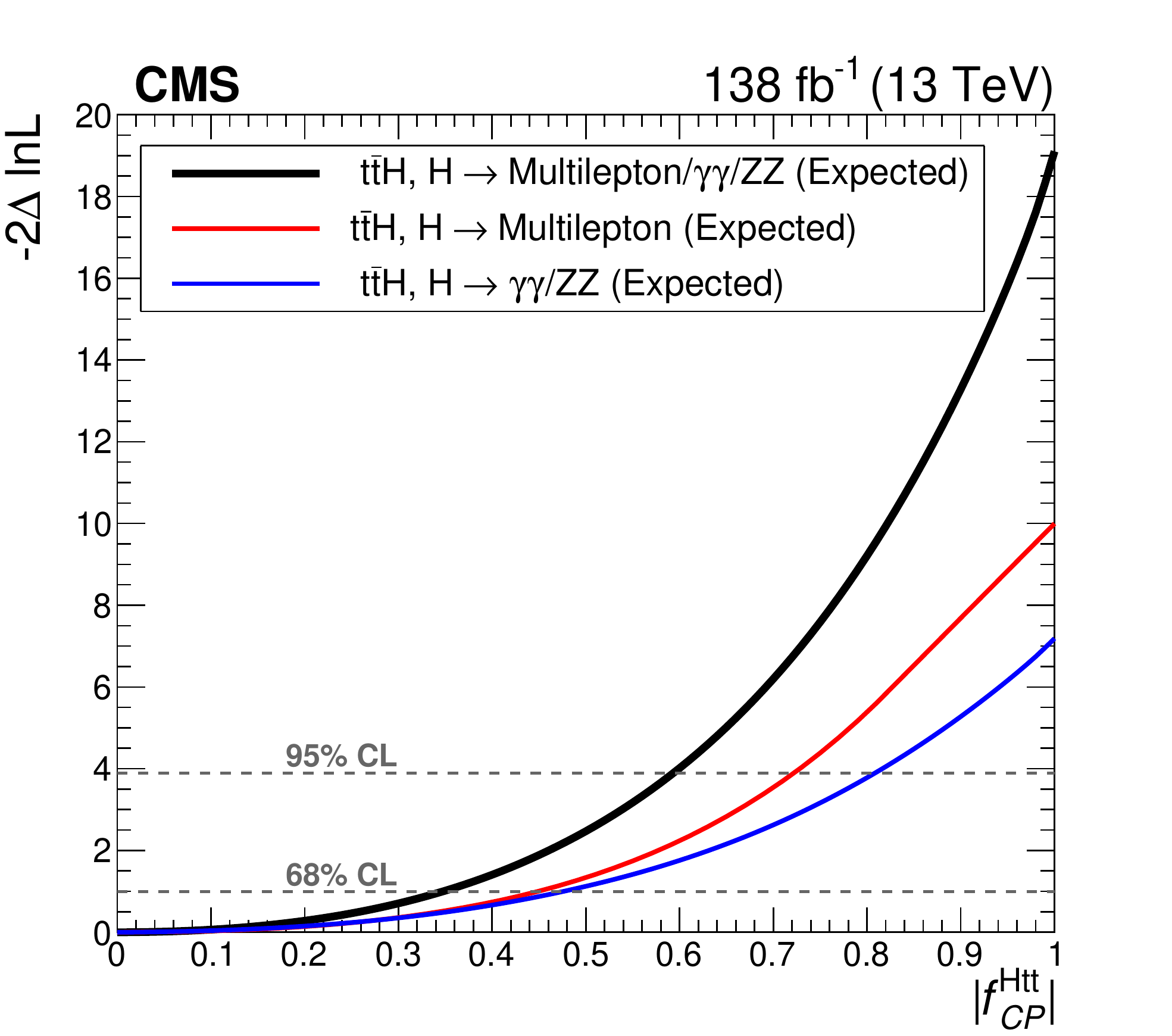}
    \includegraphics[width=0.45\textwidth]{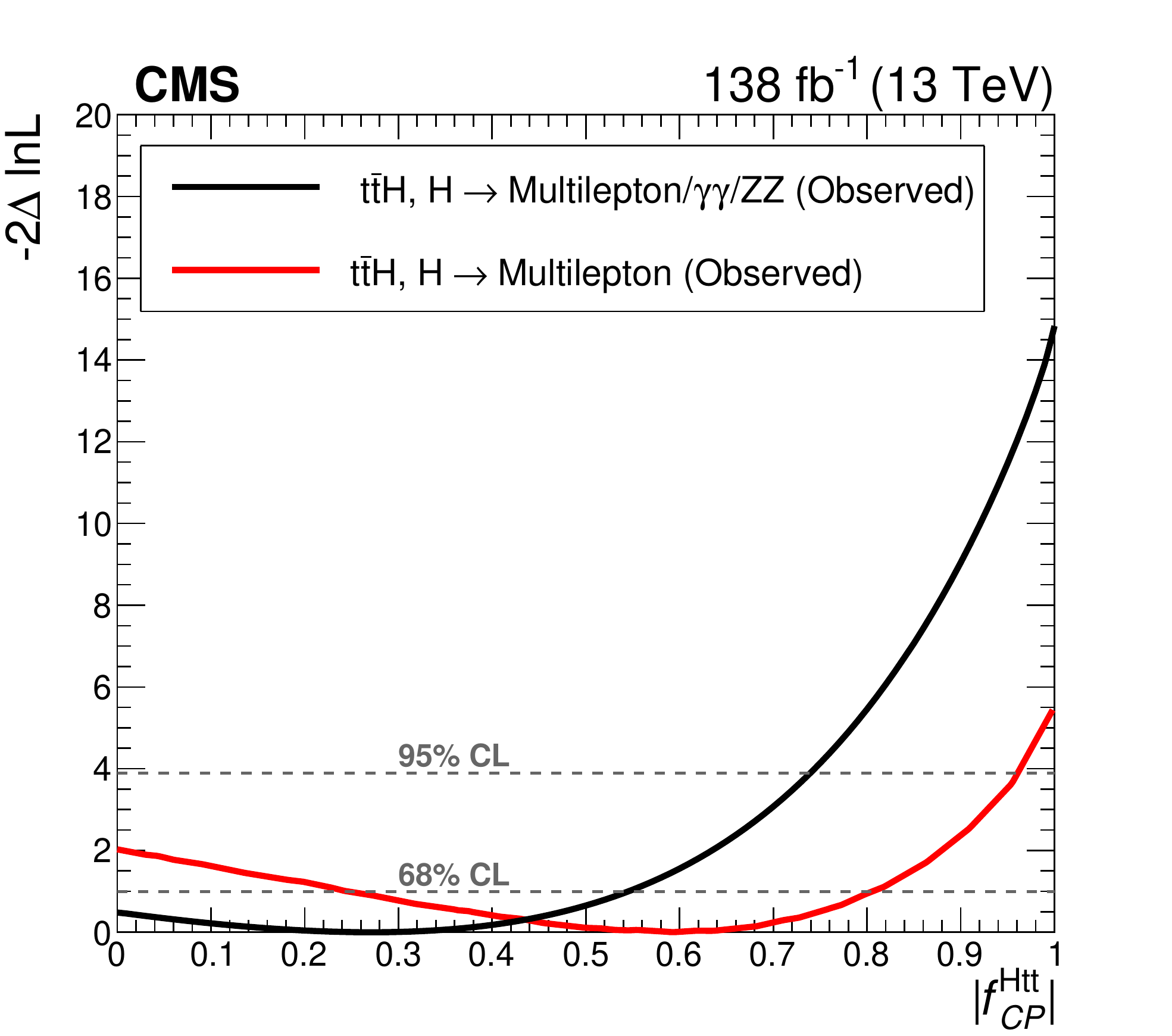}
    \caption{Likelihood scan as a function of $\fHtt$. The left plot shows the expected likelihood scan for multilepton final states, $\PH\to\PGg\PGg$, and $\PH\to\PZ\PZ$ final states, and the combination of multilepton, $\PH\to\PGg\PGg$, and $\PH\to\PZ\PZ$ final states. The right plot shows the observed likelihood scan for multilepton final states and the combination of multilepton, $\PH\to\PGg\PGg$, and $\PH\to\PZ\PZ$ final states.}
    \label{fig:fcp_exp_obs}
\end{figure}

\begin{figure}
    \centering
    \includegraphics[width=0.6\textwidth]{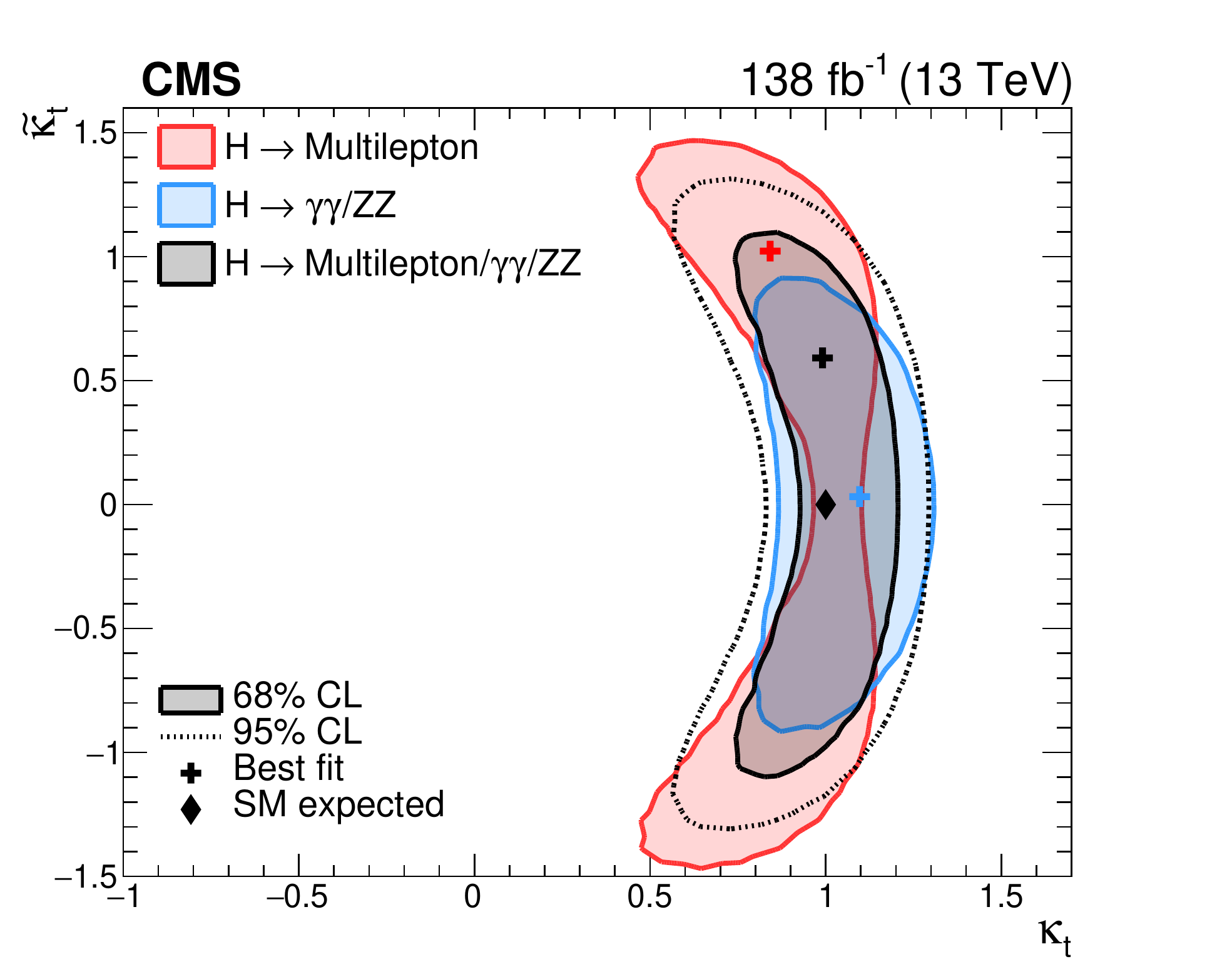}
    \caption{Likelihood scan as a function of \ckt{} and \ktt{}. Two-dimensional confidence intervals at 68\% \CL are depicted as shaded areas, for multilepton (red), the combination of $\PH\to\PGg\PGg$ and $\PH\to\PZ\PZ$ (blue), and the combination of the three channels (black). The 95\% \CL for the combination is show as a dashed line. The best fit for each is shown as a cross of the corresponding colour. The plot is symmetric with respect to the line $\ktt$=0, hence there are two points corresponding to the best fit, here we only show one for simplicity. The black diamond shows the SM expected value. The nontrivial correlation between the measurements are the source of the change in the best fit value and shape of the confidence regions. The coupling \kv{} and the \PH{} boson branching fractions are kept to their SM values.}
    \label{fig:scan_combined}
\end{figure}

The combination yields a significant improvement in sensitivity, as shown in Fig.~\ref{fig:fcp_exp_obs} (left). Confidence regions at 68 and 95\% \CL for the \ckt{}  and \ktt{} couplings are obtained using the strategy explained in this section.
Figure~\ref{fig:scan_combined} shows the likelihood scan, as a function of \ckt{} and \ktt{}, for the combination of the multilepton,
$\PH\to\PZ\PZ$, and $\PH\to\PGg\PGg$ decay modes.
 
One-dimensional confidence intervals at 68 and 95\% \CL are also obtained for \ckt{} (fixing \ktt{} to its SM value) and for \ktt{} (fixing \ckt{} to its SM value), as listed in Table~\ref{tab:CL_1d_comb}.

\begin{table}
\centering
\topcaption{One-dimensional confidence intervals at 68 and 95\% \CL for \ckt{} and \ktt{}.}
\label{tab:CL_1d_comb}
\begin{tabular}{lcc}
\hline
Parameter & 68\% \CL & 95\% \CL \\
\hline
\ckt{} & (0.96, 1.16)   & (0.86, 1.26)\\
\ktt   &  (-0.86, 0.85) & (-1.07, 1.07)\\
\hline
\end{tabular}

\end{table}

 The likelihood as a function of $\fHtt$, while profiling $\mu_{\ttH}$, is shown for the combination of the multilepton, $\PH\to\PZ\PZ$, and $\PH\to\PGg\PGg$ decay modes in Fig.~\ref{fig:fcp_exp_obs} (right).
This parameterization results in a probe for a possible fractional \CP-odd contribution, yielding a best fit value of $\fHtt =0.28$ with an interval of $\fHtt<0.55$ at 68\% \CL. The result is compatible with the SM \CP-even scenario within 68\% \CL. In addition, the scenario with $\fHtt = 1$ is excluded with 3.7 standard deviations.

\section{Summary}
\label{sec:summary}

A measurement of the charge--parity (\CP) structure of the Yukawa coupling between the Higgs (\PH) boson and top quarks at tree level, when the \PH{} boson is produced in association with one (\tH) or two (\ttH) top quarks, is presented. The measurement is based on data collected in proton-proton collisions at $\sqrt{s}=13\TeV$ by the CMS experiment at the CERN LHC, corresponding to an integrated luminosity of 138\fbinv. The analysis targets events where the \PH{} boson decays to leptons and the top quark(s) decay either leptonically or hadronically. Separation of \CP-even from \CP-odd scenarios is achieved by applying machine learning techniques to final states characterized by the presence of at least two leptons. Two-dimensional confidence regions are set on \ckt{} and \ktt{} which are respectively the \CP-even and \CP-odd top-Higgs Yukawa coupling modifiers: one-dimensional confidence intervals are also set, constraining \ckt{} to be within $(-1.09, -0.74)$ or $(0.77, 1.30)$ and \ktt{} to be within $(-1.4, 1.4)$ at 95\% confidence level (\CL).  No significant \CP-odd contribution is observed, and the corresponding fraction parameter is determined to be $\fHtt = 0.59$ with an interval of $(0.24, 0.81)$ at 68\% \CL.
The results are combined with previously published analyses covering the $\PH\to\PZ\PZ$ and $\PH\to\PGg\PGg$ decay modes. Two- and one-dimensional confidence regions are set on \ckt{} and \ktt{}, constraining \ckt{} to be within $(0.86, 1.26)$ and \ktt{} to be within $(-1.07, 1.07)$ at 95\% \CL. The possibility of a \CP-odd contribution is also investigated in the combination, yielding a best fit of $\fHtt = 0.28$ with an interval of $\fHtt<0.55$ at 68\% \CL. The results are compatible with predictions for the standard model \PH boson.

\begin{acknowledgments}
We congratulate our colleagues in the CERN accelerator departments for the excellent performance of the LHC and thank the technical and administrative staffs at CERN and at other CMS institutes for their contributions to the success of the CMS effort. In addition, we gratefully acknowledge the computing centres and personnel of the Worldwide LHC Computing Grid and other centres for delivering so effectively the computing infrastructure essential to our analyses. Finally, we acknowledge the enduring support for the construction and operation of the LHC, the CMS detector, and the supporting computing infrastructure provided by the following funding agencies: BMBWF and FWF (Austria); FNRS and FWO (Belgium); CNPq, CAPES, FAPERJ, FAPERGS, and FAPESP (Brazil); MES and BNSF (Bulgaria); CERN; CAS, MoST, and NSFC (China); MINCIENCIAS (Colombia); MSES and CSF (Croatia); RIF (Cyprus); SENESCYT (Ecuador); MoER, ERC PUT and ERDF (Estonia); Academy of Finland, MEC, and HIP (Finland); CEA and CNRS/IN2P3 (France); BMBF, DFG, and HGF (Germany); GSRI (Greece); NKFIH (Hungary); DAE and DST (India); IPM (Iran); SFI (Ireland); INFN (Italy); MSIP and NRF (Republic of Korea); MES (Latvia); LAS (Lithuania); MOE and UM (Malaysia); BUAP, CINVESTAV, CONACYT, LNS, SEP, and UASLP-FAI (Mexico); MOS (Montenegro); MBIE (New Zealand); PAEC (Pakistan); MES and NSC (Poland); FCT (Portugal); MESTD (Serbia); MCIN/AEI and PCTI (Spain); MOSTR (Sri Lanka); Swiss Funding Agencies (Switzerland); MST (Taipei); MHESI and NSTDA (Thailand); TUBITAK and TENMAK (Turkey); NASU (Ukraine); STFC (United Kingdom); DOE and NSF (USA).

\hyphenation{Rachada-pisek} Individuals have received support from the Marie-Curie programme and the European Research Council and Horizon 2020 Grant, contract Nos.\ 675440, 724704, 752730, 758316, 765710, 824093, 884104, and COST Action CA16108 (European Union); the Leventis Foundation; the Alfred P.\ Sloan Foundation; the Alexander von Humboldt Foundation; the Belgian Federal Science Policy Office; the Fonds pour la Formation \`a la Recherche dans l'Industrie et dans l'Agriculture (FRIA-Belgium); the Agentschap voor Innovatie door Wetenschap en Technologie (IWT-Belgium); the F.R.S.-FNRS and FWO (Belgium) under the ``Excellence of Science -- EOS" -- be.h project n.\ 30820817; the Beijing Municipal Science \& Technology Commission, No. Z191100007219010; the Ministry of Education, Youth and Sports (MEYS) of the Czech Republic; the Hellenic Foundation for Research and Innovation (HFRI), Project Number 2288 (Greece); the Deutsche Forschungsgemeinschaft (DFG), under Germany's Excellence Strategy -- EXC 2121 ``Quantum Universe" -- 390833306, and under project number 400140256 - GRK2497; the Hungarian Academy of Sciences, the New National Excellence Program - \'UNKP, the NKFIH research grants K 124845, K 124850, K 128713, K 128786, K 129058, K 131991, K 133046, K 138136, K 143460, K 143477, 2020-2.2.1-ED-2021-00181, and TKP2021-NKTA-64 (Hungary); the Council of Science and Industrial Research, India; the Latvian Council of Science; the Ministry of Education and Science, project no. 2022/WK/14, and the National Science Center, contracts Opus 2021/41/B/ST2/01369 and 2021/43/B/ST2/01552 (Poland); the Funda\c{c}\~ao para a Ci\^encia e a Tecnologia, grant CEECIND/01334/2018 (Portugal); the National Priorities Research Program by Qatar National Research Fund; MCIN/AEI/10.13039/501100011033, ERDF ``a way of making Europe", and the Programa Estatal de Fomento de la Investigaci{\'o}n Cient{\'i}fica y T{\'e}cnica de Excelencia Mar\'{\i}a de Maeztu, grant MDM-2017-0765 and Programa Severo Ochoa del Principado de Asturias (Spain); the Chulalongkorn Academic into Its 2nd Century Project Advancement Project, and the National Science, Research and Innovation Fund via the Program Management Unit for Human Resources \& Institutional Development, Research and Innovation, grant B05F650021 (Thailand); the Kavli Foundation; the Nvidia Corporation; the SuperMicro Corporation; the Welch Foundation, contract C-1845; and the Weston Havens Foundation (USA).
\end{acknowledgments}

\bibliography{auto_generated}
\cleardoublepage \appendix\section{The CMS Collaboration \label{app:collab}}\begin{sloppypar}\hyphenpenalty=5000\widowpenalty=500\clubpenalty=5000\input{HIG-21-006-public-authorlist.tex}\end{sloppypar}
%%% END EDITABLE REGION %%%
% skeleton_end
\end{document}

%% file: HIG-21-006-public-authorlist.tex
\cmsinstitute{Yerevan Physics Institute, Yerevan, Armenia}
{\tolerance=6000
A.~Tumasyan\cmsAuthorMark{1}\cmsorcid{0009-0000-0684-6742}
\par}
\cmsinstitute{Institut f\"{u}r Hochenergiephysik, Vienna, Austria}
{\tolerance=6000
W.~Adam\cmsorcid{0000-0001-9099-4341}, J.W.~Andrejkovic, T.~Bergauer\cmsorcid{0000-0002-5786-0293}, S.~Chatterjee\cmsorcid{0000-0003-2660-0349}, K.~Damanakis\cmsorcid{0000-0001-5389-2872}, M.~Dragicevic\cmsorcid{0000-0003-1967-6783}, A.~Escalante~Del~Valle\cmsorcid{0000-0002-9702-6359}, P.S.~Hussain\cmsorcid{0000-0002-4825-5278}, M.~Jeitler\cmsAuthorMark{2}\cmsorcid{0000-0002-5141-9560}, N.~Krammer\cmsorcid{0000-0002-0548-0985}, L.~Lechner\cmsorcid{0000-0002-3065-1141}, D.~Liko\cmsorcid{0000-0002-3380-473X}, I.~Mikulec\cmsorcid{0000-0003-0385-2746}, P.~Paulitsch, F.M.~Pitters, J.~Schieck\cmsAuthorMark{2}\cmsorcid{0000-0002-1058-8093}, R.~Sch\"{o}fbeck\cmsorcid{0000-0002-2332-8784}, D.~Schwarz\cmsorcid{0000-0002-3821-7331}, S.~Templ\cmsorcid{0000-0003-3137-5692}, W.~Waltenberger\cmsorcid{0000-0002-6215-7228}, C.-E.~Wulz\cmsAuthorMark{2}\cmsorcid{0000-0001-9226-5812}
% ...
\par}
\vskip\cmsinstskip
\dag:~Deceased\\
$^{1}$Also at Yerevan State University, Yerevan, Armenia\\
$^{2}$Also at TU Wien, Vienna, Austria\\
% ...
